\documentclass[a4paper,11pt]{article}
\pdfoutput=1
\usepackage{jheppub}
\usepackage{graphicx,slashed,hyperref,xcolor,multirow}
\usepackage{amssymb}
\usepackage[normalem]{ulem}
\usepackage[utf8]{inputenc}
\usepackage{array}
\usepackage{ctable} 
\hypersetup{
   bookmarks=true,         
   unicode=true,          
   pdftoolbar=true,        
   pdfmenubar=true,        
   pdffitwindow=false,     
   pdfstartview={FitH},    
   pdftitle={My title},    
   pdfauthor={Author},     
   pdfsubject={Subject},   
   pdfcreator={Creator},   
   pdfproducer={Producer}, 
   pdfkeywords={keyword1} {key2} {key3}, 
   pdfnewwindow=true,      
   colorlinks=true,       
   linkcolor=blue,          
   citecolor=magenta,        
   filecolor=magenta,      
   urlcolor=cyan           
}

\def\lsim{\mathrel{\rlap{\lower4pt\hbox{\hskip1pt$\sim$}}
    \raise1pt\hbox{$<$}}}         
\def\gsim{\mathrel{\rlap{\lower4pt\hbox{\hskip1pt$\sim$}}
    \raise1pt\hbox{$>$}}}         

\makeatletter

\newcommand{\fmslash}[2][0mu]{%
  \mathchoice
    {\fmsl@sh\displaystyle{#1}{#2}}%
    {\fmsl@sh\textstyle{#1}{#2}}%
    {\fmsl@sh\scriptstyle{#1}{#2}}%
    {\fmsl@sh\scriptscriptstyle{#1}{#2}}}
\newcommand{\fmsl@sh}[3]{%
  \m@th\ooalign{$\hfil#1\mkern#2/\hfil$\crcr$#1#3$}}
\makeatother

\newcommand{\beq}{\begin{equation}}
\newcommand{\eeq}{\end{equation}}
\newcommand{\bea}{\begin{eqnarray}}
\newcommand{\eea}{\end{eqnarray}}
\mathchardef\minus="002D

\usepackage{xcolor}

\def\beq{\begin{equation}}
\def\eeq{\end{equation}}
\def\bea{\begin{eqnarray}}
\def\eea{\end{eqnarray}}

\title{LHC Signals for KK Graviton from an Extended  Warped Extra Dimension}

\author[a]{Kaustubh Agashe,}
\author[a]{Majid Ekhterachian,}
\author[b]{Doojin Kim,}
\author[a]{and Deepak Sathyan}

\affiliation[a]{Maryland Center for Fundamental Physics, Department of Physics, University of Maryland,
     College Park, MD 20742, USA}
\affiliation[b]{Mitchell Institute for Fundamental Physics and Astronomy,  Department of Physics and Astronomy, Texas A\&M University, College Station, TX 77843, USA}

\emailAdd{kagashe@umd.edu}
\emailAdd{ekhtera@umd.edu}
\emailAdd{doojin.kim@tamu.edu}
\emailAdd{dsathyan@umd.edu}

\preprint{
\begin{minipage}{5cm}
\begin{flushright}
UMD-PP-020-3 \\
MI-TH-2023
 \end{flushright}
\end{minipage}
}

\abstract{
We analyze signals at the Large Hadron Collider (LHC) from production and decay of Kaluza-Klein (KK) gravitons in the context of ``extended'' warped extra-dimensional models, where the standard model (SM) Higgs and fermion fields are restricted to be in-between the usual ultraviolet/Planck brane and a $\sim O(10)$ TeV (new, ``intermediate'') brane, whereas the SM gauge fields (and gravity) propagate further down to the $\sim O( \hbox{TeV} )$ infrared brane.
Such a framework suppresses flavor violation stemming from KK particle effects, while keeping the KK gauge bosons and gravitons accessible to the LHC.
We find that the signals from KK graviton are significantly different than in the standard warped model.
This is because the usually dominant decay modes of KK gravitons into top quark, Higgs and longitudinal $W/Z$ particles are suppressed by the above spatial separation between these two sets of particles, thus other decay channels are allowed to shine themselves.
In particular, we focus on two novel decay channels of the KK graviton. 
The first one is the decay into a pair of radions, each of which decays (dominantly) into a pair of SM gluons, resulting in a resonant 4-jet final state consisting of two pairs of dijet resonance. 
On the other hand, if the radion is heavier and/or KK gluon is lighter, then the KK graviton mostly decays into a KK gluon and a SM gluon. 
The resulting KK gluon has a significant decay branching fraction into radion and SM gluon, thereby generating (again) a 4-jet signature, but with a different underlying event topology, i.e., featuring now three different resonances.
We demonstrate that the High-Luminosity LHC (HL-LHC) has sensitivity to KK graviton of (up to) $\sim 4$~TeV in both channels, whereas it is unlikely to have sensitivity in the standard dijet resonance search channel from KK graviton decay into two gluons.  
\newpage
}

\begin{document}

\maketitle

\section{Introduction}

A multitude of extensions of the standard model (SM) at the TeV scale have been proposed, in part for addressing the Planck-weak hierarchy problem of the SM and/or for furnishing a candidate for dark matter of the universe. 
With this motivation, 
searches have been performed at the Large Hadron Collider (LHC) for signals from direct
production of the associated new particles
and at lower energy flavor physics-type experiments where {\em in}direct effects (i.e., from virtual exchanges) of these particles can be manifested.
To begin with, these exercises were undertaken
for ``minimal'' incarnations in the beyond-SM (BSM) frameworks. 

The lack of conclusive new-physics signals thus far -- which results in strong constraints on the masses of the new particles -- could imply that the new particles are simply heavier than what was expected. 
Alternatively, the concrete realization of the {\em general} BSM physics could be {\em non}-minimal; indeed, even simple modifications of the vanilla model can change the signals dramatically, thus allowing the new particles to be lighter than the bounds for the minimal case simply by evading the searches in the {\em conventional} channels.
Such a situation motivates new, dedicated searches for {\em non}-standard signals. 
Overall, this situation has prompted both theorists and experimentalists to be interested in studying such variations of the standard BSM models.

\begin{figure}[tbp]
\center
\includegraphics[width=1.0\linewidth]{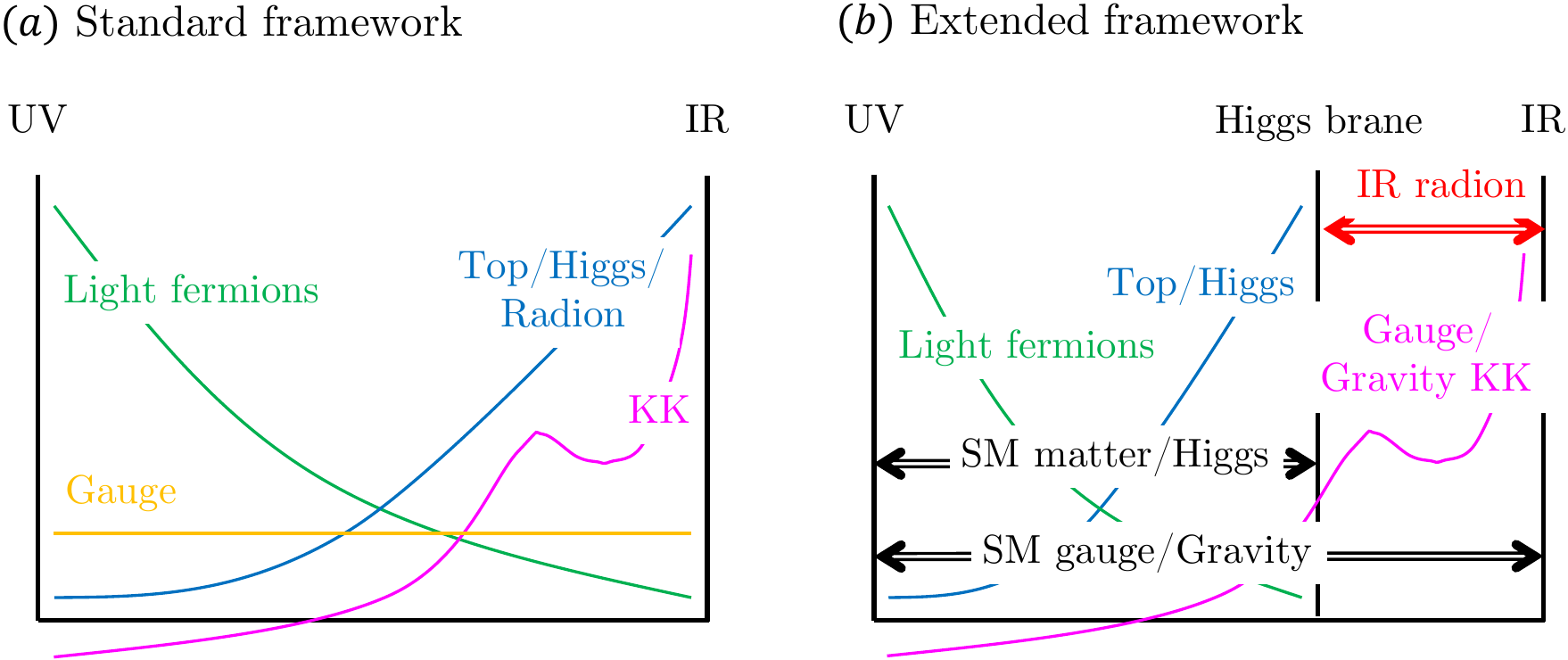}
\caption{Left: Warped extra-dimensional model with SM fields in the same bulk (standard framework). Schematic shapes of extra-dimensional wavefunctions for various particles (zero mode SM fermions and gauge particles, a radion, and a generic KK mode) are shown. 
Right: Warped extra-dimensional model with {\em all} SM gauge fields and gravity in the full bulk, but matter and Higgs fields in subspace (extended framework). Schematic shapes of extra-dimensional wavefunctions for various particles (zero mode SM fermions and gauge bosons, an IR radion, and gauge/gravity KK modes) are shown.
The double-lined arrows indicate that the associated fields propagate within the marked range in the extra dimension.}
\label{fig:standard}
\end{figure}

A nice illustration of the general theme outlined above is provided by the framework of a warped extra dimension (that is our focus here) into which SM fields propagate: for reviews, see \cite{Sundrum:2005jf,Csaki:2005vy,Davoudiasl:2009cd,Gherghetta:2010cj, Ponton:2012bi}: 
this paradigm can address the Planck-weak and flavor hierarchy problems of the SM.
In the minimal scenario (henceforth called ``standard warped model''), there are 
two branes: the ultra-violet (UV)/Planck brane near which the light SM fermions are localized, and the infra-red (IR)/TeV brane where the SM Higgs field and the top quark are situated, as schematically shown in figure~\ref{fig:standard}($a$).
By contrast, the SM gauge bosons are essentially delocalized in the extra dimension.
The new particles in this setup are the Kaluza-Klein (KK) excitations of SM particles (and graviton) and radion (which is roughly the modulus corresponding to fluctuation in size of extra dimension): they {\em all} reside near the IR/TeV brane and their masses are roughly given by the IR brane scale, i.e., $O (\hbox{TeV} )$.

The resulting (extensively studied) signals for this scenario involve heavy ($\gtrsim$
TeV) resonances such as the KK modes and radion decaying into a pair of SM particles, preferably the heavy ones such as top quark and Higgs particles, including longitudinal $Z/W$, as all three particles involved in the relevant couplings are localized near the IR brane (i.e., a ``nearest neighbor'' effect).
However, the null results from the LHC suggest that these new particles are heavier than a few TeV
(see, for example, \cite{Sirunyan:2018ryr,Aaboud:2019roo}); in fact, flavor/CP violation constraints
(without extra symmetries) already required $\gtrsim O(10)$ TeV scale for their masses \cite{Csaki:2008zd,Blanke:2008zb,Bauer:2009cf,KerenZur:2012fr} even before the LHC was turned on,
whereas, electroweak precision tests can be satisfied even for a few TeV KK scale \cite{Agashe:2003zs,Agashe:2006at,Carena:2006bn,Carena:2007ua,Delaunay:2010dw}.

The above status of the standard warped model (in part) has led to the proposal of a (literal) {\em extension} in ref.~\cite{Agashe:2016rle} as follows. 
Imagine that 
\begin{itemize}
\item
\noindent
the SM Higgs and fermion fields only
live from the UV/Planck brane to a $\sim O(10)$ TeV brane (henceforth called
``Higgs brane''), thereby satisfying flavor/CP constraints. However, SM gauge fields (and gravity) 
``continue'' to propagate beyond this Higgs brane to a few TeV IR brane: see figure~\ref{fig:standard}($b$). 

\end{itemize}
That various fields occupy different regions of the bulk -- as above -- also seems plausible from a ``UV'', for example, string theory, viewpoint.
Moreover, 
the presence of extra brane(s) does not have to be taken literally.
We can consider such a setup simply as a phenomenological way to model an IR region of the warped extra dimension which has more structure than  just  one,  featureless  (“bare”)  brane  (often  called  “hard”  wall);  in  this  sense  it  is somewhat similar
in spirit to adding IR-brane-localized kinetic terms for bulk 
fields \cite{Davoudiasl:2002ua,Carena:2002dz,Carena:2004zn,Davoudiasl:2003zt} or a ``soft'' wall (see, for example, \cite{Carmona:2011ib} and references therein), i.e., metric deviating from pure AdS in the IR, but with {\em different} actual physics than these previous scenarios.

Finally, as is well-known (for a review, see \cite{Gherghetta:2010cj}), by the AdS/CFT correspondence, the standard
warped model is conjectured to be a weakly-coupled dual description of a
(purely) 4D scenario with the SM Higgs being a composite arising from a new strong dynamics (while rest of the SM particles are admixtures
of composites and elementary/external states).
As discussed at length in ref.~\cite{Agashe:2016rle}, the extended warped framework would then correspond to Higgs
compositeness primarily
at $\sim O(10)$ TeV, but with (lighter) ``vestiges'' of this composite sector appearing at a (slightly) smaller mass scale, $O$(TeV), in the form of spin-1, spin-2, and spin-0 resonances, i.e., KK gauge bosons, KK gravitons, and radion.

The upshot is that with such a modification of the standard warped model, the gauge and graviton KK modes (and radion) -- being $O ( \hbox{TeV} )$ in mass -- are potentially within reach of the LHC.
However, the KK fermions are likely {\em not} accessible (directly) at the LHC, since their masses are dictated by the Higgs brane scale, i.e.,  
are of $O(10)$ TeV.
Remarkably, the signals from gauge/graviton KK and radion production are significantly modified relative to
the standard warped model. Namely, 
\begin{itemize}
\item
\noindent
the usually dominant decay modes of KK particles/radion into top quark/Higgs particles
are rendered small due to their ``sequestering''
from these new particles: again, these couplings are all governed by overlap of profiles in the bulk: see figure~\ref{fig:standard}($b$) vs. figure~\ref{fig:standard}($a$); 
\end{itemize}
this permits 
\begin{itemize}
\item
\noindent
other, ``pre-existing'' 
modes -- hitherto swamped by decays 
into top quark/Higgs particles -- to emerge
as the leading ones.

\end{itemize}
Such signals have been studied for {\it gauge} KK particles in refs.~\cite{Agashe:2016kfr,Agashe:2017wss,Agashe:2018leo}.

Here, we continue and expand this program by analyzing LHC signals
from KK {\em graviton} production and decay in this new scenario (called ``extended warped model'' from now on); such a study is motivated on 
various fronts as follows: 
\begin{itemize}

\item
\noindent
Obviously, KK graviton has a {\it different} spin of 2 than gauge KK of spin 1 so that in general angular distributions of the decay products can indeed distinguish the two particles even if
final state particles are the same. 
We will show that this angular distribution of the decay products of the KK graviton, a characteristic feature of its spin-2 nature, enables discrimination of its signal from SM background as well, which is crucial for expediting its discovery.

\item Indeed, KK graviton is a tell-tale sign of extra dimensions\footnote{Of course, in the warped case, this scenario is dual to SM particles being (partially) composites of new (purely) 4D strong dynamics.} vs. spin-1 gauge KK particles which could also arise in purely 4D extensions of the SM.

\item Moreover, the final state from KK graviton decay in the ``updated'' (i.e., in the extended warped model) dominant mode is also itself different than that of the gauge KK decay channel, cf. standard warped model where the dominant decay modes for the KK gauge and graviton particles are similar (as indicated above), i.e., both give a pair of heavy SM particles in the usual case.Hence, the ``extended'' searches proposed in refs.~\cite{Agashe:2016kfr,Agashe:2017wss,Agashe:2018leo} for gauge KK modes might {\em not} apply to the KK graviton, cf. standard warped model where 
the same channel can be sensitive to both KK gauge and graviton modes.

\item Finally, the topology of the final state from the KK graviton decay in the new dominant channel has {\em not} really been experimentally searched for thus far, say, even in the context of other models, so that existing experimental analyses will not be efficient at catching this signal.
\end{itemize}
Overall, we emphasize that new {\it dedicated} searches might be needed for effective discovery at the LHC of the KK modes (and radion) in the extended warped model, that too separate ones for gauge and graviton.
Specifically, we study two {\em signal regions} (SRs) associated with the KK graviton in this model.

\medskip

\noindent {\bf SR I}: If the radion mass is smaller than one-half of KK graviton mass,\footnote{Generically radion is expected to be lighter than KK graviton, but here we need an extra assumption.}
then KK graviton has a sizable decay branching ratio (BR) into a pair of radions since the relevant coupling has roughly the largest possible size due to the fact that all three (KK graviton and two radions) profiles are pinned to the IR brane, whereas the competing modes into top quark/Higgs particles are suppressed as already mentioned previously.\footnote{\label{ft:footnote}Actually, although decays of KK graviton to gauge modes, whether one KK and one SM or both SM, involve couplings which are smaller than in the decay to radions, the gauge final states' larger number of degrees of freedom (including
spin and color) can compensate, making them comparable to (or even larger than) the decay to a pair of radions (again, based simply on non-kinematic factors). 
So, in order to make BR for decays of KK graviton into a radion pair significant, we can (kinematically) suppress at least the decay into gauge KK particle(s) by choosing the mass gap between KK graviton and gauge KK modes small enough (even though latter is generically expected to be lighter than former).}
\bea
\hbox{KK graviton (spin-2)} & \rightarrow & \hbox{2 radions (spin-0)}  
\eea
Note that even in the standard warped model, the decay of KK graviton into a pair of radions occurs; this is a good place to re-emphasize that all of the ``new'' signals from KK graviton that we discuss here for the extended warped model (similarly, for KK gauge modes done earlier) originate from couplings that are {\em inherent} even to the standard warped model
(and with similar size as in the extended version), i.e., they are not ``engineered'' in order to generate these signals.  
It is merely that (for various reasons outlined below) the effects of these couplings on decays of KK modes were not apparent in the standard warped model.

\begin{figure}[t]
    \centering
    \includegraphics[width=15.cm]{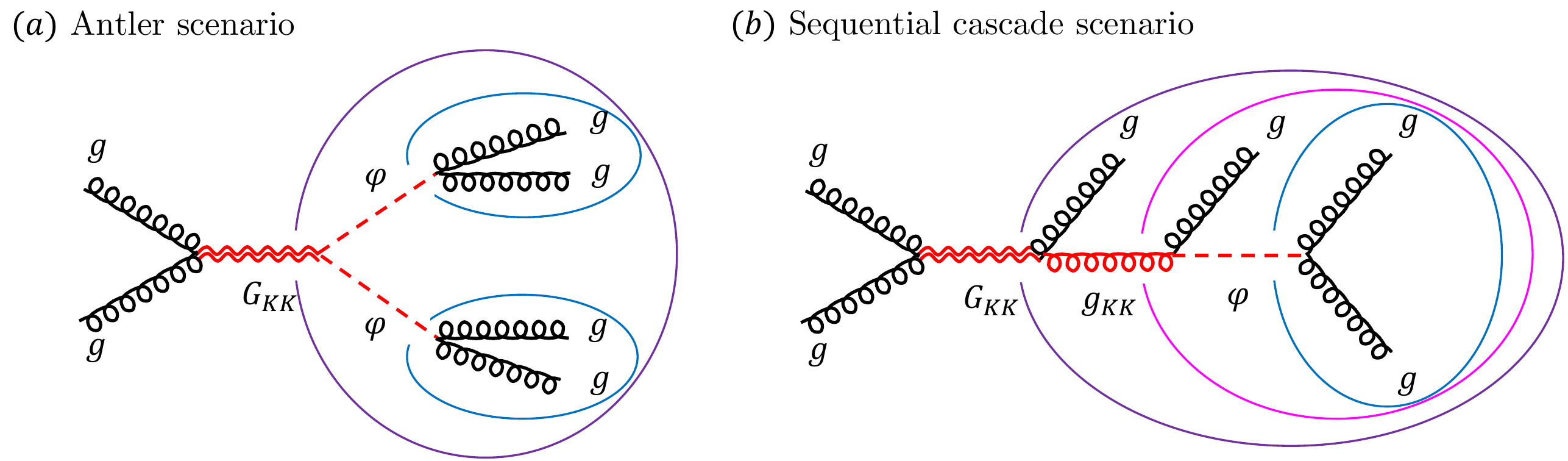}
    \caption{Two possible event topologies arising at the LHC from KK graviton production/decay that we study in the extended warped model under consideration. Final-state gluons forming KK graviton $G_{\rm KK}$, KK gluon $g_{\rm KK}$, and radion $\varphi$ are grouped by purple, pink, and blue circles, respectively.
    }
    \label{fig:evetopo}
\end{figure}

In turn, in the extended warped model, each {\em radion} from the above KK graviton decay mostly decays into two SM gluons.\footnote{cf.~in the standard warped model, where pair of top quarks/Higgs particles -- again, just like for KK modes -- is favored.}
\bea
\hbox{radion (spin-0)} & \rightarrow & \hbox{2 SM gluons}  
\eea
This is another example of the general point made above, namely, the decay mode of radion into SM gluons exists (again, with similar coupling) even in the standard warped model, but is overwhelmed by that into heavy SM particles simply based on comparison of the respective couplings, in turn, determined by overlap of profiles. 
In other words, the former coupling involves two delocalized profiles (for SM gluons) and one localized near the IR brane (radion) vs. all three being concentrated near the IR brane for the latter.
We see that the impact of the suppression of the couplings of radion to top quark/Higgs particles
in the extended warped model (again due to the ``splitting'' of these particles from each other in the extra dimension) is to convert what was a {\em sub}-dominant channel (i.e., radion decay into SM gluons) in the standard case into the leading one.
\begin{itemize}

\item
\noindent
We then get a 4-jet resonance, constituted by two identical pairs of dijet resonances, i.e., 
the so-called 
``antler'' topology \cite{Han:2009ss, Han:2012nm}: see figure~\ref{fig:evetopo}($a$).
We henceforth denote this decay channel by {\it radion channel}.

\end{itemize}

It is clear that the decay BR of KK graviton into radions is actually comparable\footnote{Up to degree-of-freedom (including spin and color) factors.}
to that into top/Higgs even for the {\em standard} warped model, again because all these particles -- KK particles, top quark/Higgs field, radion -- are localized near the IR/TeV brane.
However, the decay of KK graviton to radions followed by the decay of each radion into top quark/Higgs particles has not been really studied in the literature thus far for the standard warped model, presumably because the ``di-(heavy) SM'' modes are easier to analyze.\footnote{Reference \cite{Dillon:2016tqp} did study this decay channel, assuming decay to SM fermions/Higgs is suppressed, but focused mostly on the case of a very light radion (below $2 \; m_{ \pi }$, forbidding decays into 2 SM gluons, i.e., dijet) so that it decays into a pair of photons which merge and give photon-jets.}
By contrast, in the extended warped model, the decay mode into a pair of radion is singled out (from amongst these decay modes) as the leading channel because of the suppression mentioned above of erstwhile (i.e., in the standard setting) {\em co}-dominant modes.

Also, curiously enough (to the best of our knowledge), 
\begin{itemize}

\item
\noindent
there are current searches for pairs of dijet resonances, but without requiring an {\em overall} resonance as these were motivated by, for example, pair production of superpartners of top quark, followed by their decay -- via $R$-parity breaking couplings -- into two quarks.

\end{itemize}
In fact, motivated by an excess seen by CMS in this topology (see, for example, a 4-jet event shown in Figure~3 of \cite{Sirunyan:2019vgj}), it seems that recently a search along these lines has been commissioned: our model then can provide a motivation for it.
Other models also possess a 4-jet signal with a similar topology, see, for example, discussion and references in \cite{Dobrescu:2018psr}; thus, our analysis strategy can be adapted for studying these signals as well. 

\medskip

\noindent {\bf SR II}: On the other hand, if the mass of radion is greater than (or close to) one-half the mass of KK graviton, i.e., the decay of KK graviton into a pair of radions is kinematically forbidden (or simply phase-space suppressed), then the decay of KK graviton into gauge KK plus the corresponding SM gauge boson can be the leading channel, with gluon dominating here.
\bea
\hbox{KK graviton (spin-2)} & \rightarrow & \hbox{KK gluon (spin-1)} + \hbox{SM gluon} \label{eq:firststep}  
\eea
Note that gauge KK mode is generically expected to have smaller mass than KK graviton; here we 
assume that the KK gluon is sufficiently lighter, such that phase-space suppression in this decay is only mild.
Actually, as noted in footnote~\ref{ft:footnote}, due to an enhancement by the number of degrees of freedom (d.o.f.), the BR of this channel (provided that it is not phase-space suppressed) can be competitive to or even larger than that of the decay into a pair of radions, in spite of the latter channel having larger coupling and even if it is kinematically unsuppressed.

Once again, the above channel is already present even in the standard setting.
However, in that case, this decay was swamped by that into top quark/Higgs particles
(cf.~above case of decay into a pair of radions which was similar in rate to top quark/Higgs particles): again, it is all due to the profiles, i.e.,
the KK graviton-KK gauge boson-SM gauge boson coupling is made up of two profiles peaked near IR brane (KK graviton and KK gauge boson) and one delocalized (SM gauge boson), whereas the KK graviton-heavy SM pair coupling involves all three particles inhabiting the IR region, i.e., the former has the next-to-largest possible size vs. largest for the latter.\footnote{The coupling to two SM gauge bosons is even smaller, originating from the fact that two of the profiles are now delocalized.}
As usual, it is the damping of
couplings of KK graviton modes to top quark/Higgs particles (i.e., the ``usual suspects'' for decay modes) in the extended warped model -- and kinematic suppression of the decay into a radion pair (i.e., by choice of spectrum) or simply due to its smaller d.o.f.~(again, this is the only coupling which could have the maximum size) -- which unveils the usually ``next-in-line'' novel decay channel.\footnote{Here,
we consider the ``minimal'' version of the extended warped model, where only SM gauge fields propagate in the ``extra'' bulk
and there are no new fields localized on the IR brane. 
In this case, there are no (purely) BSM light particles -- except for radion -- for the KK gravitons/gauge bosons to decay into.
On the other hand, in models with extended bulk gauge symmetries in the extra bulk, which are broken down to the SM gauge group on the IR brane, there can be light $A_5$ modes into which KK gravitons/gauge bosons can decay: see, for example, Section 2.4 of ref.~\cite{Agashe:2016rle}.
LHC signals along these lines (in particular, giving 4-jet final state made up of 2 pairs of resonant jets) have been studied in the context of the 4D-dual of the extra bulk, i.e., vector-like confinement~\cite{Kilic:2008pm,Kilic:2008ub,Kilic:2009mi,Kilic:2010et}.} 

Now, the KK {\em gluon} (from the KK graviton decay) itself can decay into light quarks
($q \bar{q}$)
in a flavor-universal manner (again because top quark-dominance of the standard warped model no longer holds), resulting in a net three-jet final state from KK graviton production and decay.
It is amusing that this will ``mimic'' the final state from direct production of KK gluon itself, followed by a different decay route, e.g., into radion plus a SM gluon
with radion (in turn) decaying into two SM gluons, as already studied in ref.~\cite{Agashe:2016kfr}.
Namely, both direct KK gluon production followed by its decay into radion and that of KK graviton in the above-indicated mode feature a dijet resonance and a trijet resonance! 
So, a more dedicated analysis (for example, involving angular distributions due to different spins of the resonances) would have to be undertaken in order to distinguish the two possibilities for explaining such a signal (once it is observed).

Since the above event topology has already been studied (albeit in the context of KK gluon signal), here we instead focus on the case of KK graviton decaying mostly to a KK gluon and a SM gluon, but with KK gluon also decaying into the above-mentioned extended channel (which has comparable BR to $q\bar{q}$), namely, a radion (which itself gives two SM gluons) and a SM gluon: 
\bea
\hbox{KK gluon (spin-1)} & \rightarrow & \hbox{radion (spin-0)} + \hbox{SM gluon}, \nonumber \\
\hbox{radion (spin-0)} & \rightarrow & \hbox{2 SM gluons}. \label{eq:secondstep}
\eea
Again, this decay mode of KK gluon -- by itself, i.e., from its direct production -- 
was analyzed in ref.~\cite{Agashe:2016kfr}.  
So, considering \eqref{eq:firststep} and \eqref{eq:secondstep} together, once again, 
\begin{itemize}

\item
\noindent
we get a 4-jet resonance, but now crucially via an event topology [a ``sequential cascade'': see figure~\ref{fig:evetopo}($b$)] which is different from the earlier one (antler), i.e., encompassing two intermediary new particles (on-shell), thus a total (for the entire process) of three resonances with 
{\em all}
different masses, namely, KK graviton, KK gluon, and radion, 

\end{itemize}
while there are only {\em two} 
different resonance masses (albeit still three invariant mass bumps)
in the previous case of KK graviton decaying into two radions.
In order to distinguish it from the radion channel, we henceforth call the associated channel {\it KK gluon channel} as a KK gluon appears in the decay process. 
Indeed, discovering such a ``layered'' signal will nicely illuminate the underlying structure of the model, as also pointed out in the literature in the context of other models, e.g., photon cascade decay in the standard warped model~\cite{Agashe:2014wba} and heavy-neutrino signal in the left-right symmetric SM~\cite{Dev:2015kca}. 

\medskip 

Regarding the signal processes associated with the two SRs described thus far, it is interesting to find that 
\begin{itemize}

\item[($i$)]
(in general) with a heavy parent decaying into 4-body (in this specific case, 4-jets) final state, assuming only 2-body decays at each step of the process
(since if kinematically open, the 2-body decays likely dominate over $\geq3$-body decays), 
the only decay topologies possible are the above-mentioned antler \cite{Han:2009ss, Han:2012nm} and sequential (three-step) cascade. Thus, remarkably,
\item[($ii$)]
this model features both these allowed structures from KK graviton decay into a pair of radions and (instead) into KK gluon (followed by its decay
into radion), respectively.
\end{itemize}

A couple of comments follow regarding the signal processes under consideration. 
First, a similar cascade decay of KK graviton was studied in ref.~\cite{Agashe:2014wba} in the context of the standard warped model, i.e.,
KK graviton decaying into a SM photon (so that we get a clean, i.e., SM background-suppressed, signal) and a KK photon. 
However, as already mentioned, this decay channel was subdominant in that case 
vs.~such a cascade decay mode being dominant here in the extended model.
Moreover, decay of the KK photon itself in the study in ref.~\cite{Agashe:2014wba} was mostly into top quark/Higgs particles, whereas here the decay being either equally into a pair of all (appropriately) charged SM particles (di-SM) or the novel channel of a SM gauge boson plus a radion. 
So, the final state itself (i.e., not just the rate) could be different in the two cases: once again, we can have a dramatic departure from the usual situation.
Second, we can replace some of the gluons (KK or SM as appropriate) by electroweak (EW) gauge bosons in the above two novel KK graviton decay channels, at the price of reducing the signal by the ratio of powers of SM gauge couplings (and degrees of freedom of gluon vs.~EW gauge bosons): note, however, that corresponding SM background will also be smaller, hence there is a chance of this channel also being relevant.

Note that both among experimentalists and theorists, there is actually growing interest in going beyond the intensively studied paradigm of a heavy resonance decaying into two SM particles, i.e., in considering a (lighter) BSM resonance among the decay products of the heavy resonance \cite{Kim:2019rhy}. 
Our work here (and earlier papers on gauge KK signals) is in line with this effort; in particular, gauge KK decays into radion fit into group II/table 2 of \cite{Kim:2019rhy}, whereas the KK graviton decay into a pair of radions belongs to group Y/table 5, with
KK graviton decay into KK gluon falling into group IV/table 4.

Here is a summary of our key results. For SR I, 
we find several benchmark points which satisfy current bounds on the various new particles, while giving 
$4\sigma-5\sigma$ (statistical) significance 
(i.e., possibly discovery)
using the 4-jet antler topology. 
Moving onto SR II,
we show a few benchmark points where the 4-jet sequential cascade topology
gives $\sim 2.5\sigma -3\sigma$ (statistical) significance (i.e., potential evidence for new physics).
On the other hand, the KK graviton decay into two SM gluons, which gives a dijet final state (i.e., a standard search), is less sensitive than the antler and the sequential cascade topologies, i.e., in both SRs of parameter space.
Overall, the first discovery of KK graviton itself is likely to happen via these new channel(s), further motivating these searches, i.e., not simply because they are characterized by novel topologies.

An outline for the rest of the paper is as follows. 
We begin in section \ref{sec:model} with a brief description of the model, including 
tabulating the couplings relevant for the LHC signals that we study.
This is followed in section \ref{sec:overview} by an overview of the LHC signals; in particular, we select some benchmark points satisfying the current constraints on the parameter space of the model.
Our event simulation and selection criteria are discussed in section~\ref{sec:simulation}.
Section \ref{sec:analysis} then contains the
results of our analysis of the LHC signals.
Finally, in section \ref{sec:conclude}, we provide conclusions of our work and an outlook.

\section{Review of the Extended Warped Extra-Dimensional Model}
\label{sec:model}

The model associated with our signal processes has already been outlined in the introductory section [see figure~\ref{fig:standard}($b$)] and has been discussed in detail in the theory and phenomenology refs.~\cite{Agashe:2016rle,Agashe:2016kfr,Agashe:2017wss,Agashe:2018leo}, so we 
shall only provide a brief description here.
As usual, the SM particles are identified as zero-modes of corresponding 5D fields. 
The 5D fermion (and possibly Higgs) fields reside in-between the UV brane and an ``intermediate'' brane with scale $\sim O(10)$ TeV, where the SM Higgs field and top quark are localized (thus it will often be referred to as the ``Higgs'' brane).
The light SM fermions have profiles which are peaked near the UV brane.
By contrast, the 5D gauge fields and gravity propagate in the above ``common'' bulk, plus in an ``extended'' bulk down to the final IR brane at $\sim O ( \hbox{TeV} )$, i.e., which is only modestly (but still significantly for phenomenology) separated from the Higgs brane.
The SM gauge bosons are delocalized, i.e., have a flat profile, in the entire extra dimension.
For the sake of completeness, we mention that the 4D graviton profile is (highly) peaked near the UV brane, but it will not play any role here.

The 
new particles in this model are the Kaluza-Klein excitations of SM fields, (the lightest ones of) which are localized near the respective ``IR'' branes, i.e., KK fermions at the Higgs brane and KK gauge/graviton near the final IR brane.
The model also features another set of new particles, i.e., the two radion fields, which are the moduli associated with changes in inter-brane separations (UV-Higgs and Higgs-final IR): these are localized near the respective IR ends, i.e., Higgs and final IR branes.
The going-rate for the masses of these new particles is the scale of the relevant IR brane. 
In this paper on signals at the LHC, therefore, we will neglect the heavier radion as well as KK fermions since they have masses of $ \sim O(10)$ TeV, i.e., they are inaccessible directly at the LHC.
Instead, we 
focus on gauge/graviton KK particles and the lighter radion of mass $\sim O ( \hbox{TeV} )$.

Remarkably, as is well-known, as per the AdS/CFT correspondence as applied to warped (i.e., finite AdS$_5$) models (for a review, see \cite{Gherghetta:2010cj}), this framework is conjectured to be dual to a purely 4D scenario where the SM Higgs field is a composite of new strong dynamics.
The rest of SM fields (which are again the zero-modes of a 5D model) are partially composite, i.e., admixtures of elementary particles (i.e., external to the strong dynamics) and composites of the strong sector.
The tower of KK particles of the 5D model is then roughly (modulo the above mixing) dual to that of heavier bound states in the (purely) 4D framework.
Indeed, this dual picture (in part) inspires a simplified 4D approach  \cite{Contino:2006nn} which is convenient for phenomenological studies and which we will employ here.
Namely, we have two ``sites'': elementary and composite (matching the above picture), with the particles at these sites mixing with each other; the simplification with respect to the full model being that only Higgs and the first level of bound states are kept on the composite site (i.e., even heavier bound states are dropped), which corresponds (approximately) to restricting to only the zero and the first KK mode of the 5D model.

In particular, as already indicated in the introduction, we focus here on the following two sets of signals stemming from production of KK graviton (denoted by $G_{ \rm KK }$) involving decays into KK gluon ($g_{ \rm KK }$) and/or radion ($\varphi$): 
see figures~\ref{fig:standard}($a$) and \ref{fig:standard}($b$).
\begin{eqnarray}
  && G_{\rm KK}\to \varphi+\varphi,~\varphi \to g+g  \label{eq:SR1}\\
  && G_{\rm KK} \to g_{\rm KK}+g,~g_{\rm KK} \to \varphi + g,~\varphi \to g+g \label{eq:SRII} 
\end{eqnarray}
Here $g$ denotes the SM gluon. 
The couplings and decay widths relevant for these signals are summarized below: for a detailed  derivation, the reader is referred to \cite{Contino:2006nn} for application of the two-site picture to the standard warped model and \cite{Agashe:2016rle,Agashe:2016kfr,Agashe:2017wss,Agashe:2018leo} for the extended warped model.

\begin{figure}[t]
    \centering
    \includegraphics[width=8cm]{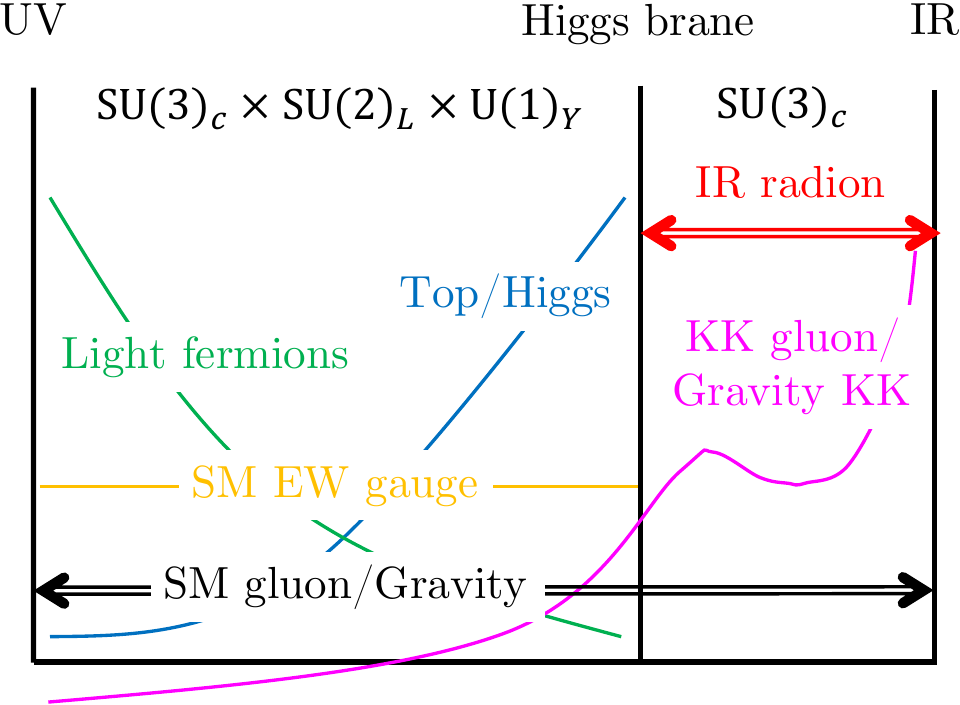}
    \caption{The model under consideration in this study that is a variation of the extended warped model framework depicted in figure~\ref{fig:standard}($b$). Only the gluon field out of the SM gauge fields is allowed to propagate in the extended bulk. }
    \label{fig:effext}
\end{figure}

For simplicity, we assume that {\em only} 
gluon field propagates in the extended bulk: see figure~\ref{fig:effext}, i.e.,
EW gauge fields are restricted from Planck brane to the Higgs brane.
Thus, KK EW gauge bosons are too heavy, i.e., of $\sim O(10)$ TeV, to be produced at the LHC.
Also, radion decays to SM EW gauge bosons can then be neglected, once again due to the spatial separation between
these two sets of particles.

\paragraph*{($a$) Radion:} Couplings of radion relevant for its production and decay are given by~\cite{Agashe:2016rle,Agashe:2016kfr,Agashe:2017wss,Agashe:2018leo}
\bea
\mathcal L^{\rm radion}_{ \rm extended \; warped } &\supset& \left(  -\frac{1}{4}\frac{g_{\rm grav}}{g_{A_{\rm KK}}^2} g_{A}^2  A_{\mu\nu}A^{\mu\nu} \right) \frac{\varphi}{m^{ (A) }_{\rm KK}}\,,
\eea
where KK graviton coupling is denoted by $g_{\rm grav}$ and $m^{ (A) }_{\rm KK}$ stands for the mass of the associated KK gauge boson.
We denote KK gauge bosons (appearing below) and SM gauge bosons collectively by $A_{\rm KK}$  and $A$, while $A^{\mu\nu}$ and $A_{\rm KK}^{\mu\nu}$ are their usual field strength tensors, and $g_A$ and $g_{A_{\rm KK}}$ parameterize their gauge couplings, respectively.
Note that after all, radion is part of the 5D gravitational field so that
it is the KK graviton 
coupling that determines the radion coupling also.\footnote{However, for simplicity and in order to
match the convention of refs.~\cite{Agashe:2016kfr,Agashe:2017wss,Agashe:2018leo}, we take the mass scale 
suppressing the radion coupling to gauge particles be that of the corresponding, i.e., gauge, 
KK mode.}
One can then easily find that the radion decay width in the limit of $m_\varphi \gg m_A$ is given by
\bea
\Gamma(\varphi \to A A ) & \approx & 
N_A \left( g_{\rm grav}\frac{g_{A}^2}{ g_{A_{\rm KK}}^2} \right )^2\left(\frac{ m_{\varphi} }{ m^{ (A) }_{\rm KK} }\right )^2 \frac{m_\varphi}{64 \pi}\,,
\label{eq:Radion_BR}
\eea
where $N_A$ is the number of degrees of freedom (not counting polarization) of SM gauge boson: 8 for gluon, 2 for $W$, and 1 for $\gamma$ and $Z$. 

\paragraph*{($b$) KK gauge bosons:} Couplings of KK gauge boson relevant for its production and decay are (see \cite{Agashe:2016rle,Agashe:2016kfr,Agashe:2017wss,Agashe:2018leo} for more details)
\bea
\mathcal L^{\rm gauge}_{ \rm extended \; warped } &\supset& \frac{g^2_A}{g_{A_{\rm KK}}} A_{\rm KK} ^\mu J_{A\mu} 
+ \epsilon \frac{g_{\rm grav}}{g_{A_{\rm KK}}} g_{A} A_{\mu\nu}A^{\mu\nu}_{\rm KK} \left( \frac{\varphi}{m^{ (A) }_{\rm KK}}  \right)\,,
\eea
where $J_{A\mu}$ is the current made of SM fields (fermions, gauge bosons and the Higgs)\footnote{However, for simplicity, we take the intermediate/Higgs brane (where the top quark/Higgs boson, including longitudinal $W/Z$, are localized) to be located far away from the IR brane 
so that the couplings of top quark/Higgs particles to gauge KK are approximately the same as that of the light fermions which are localized all the way on the UV brane.\label{ft:footnote}} associated with the SM gauge boson $A$. 
Also, the KK gauge boson-SM gauge boson-radion coupling arises after radius stabilization, thus depends on $\epsilon \sim \ln \left( \dfrac{ m^{ (f) }_{ \rm KK } }{ m^{ (G) }_{ \rm KK } } \right)$ with $m^{ (f) }_{ \rm KK }$ denoting the KK fermion mass (roughly
the
scale of Higgs brane).

In particular, the main decay channels for the KK gluon are radion (plus gluon) channel, dijet, and di-top channels, and we give the partial decay widths for the first two:
\bea
\Gamma ( g_{\rm KK} \to \varphi g) & \approx  &  \left( \epsilon  g_{\rm grav} \frac{g_{g}}{ g_{g_{\rm KK}}} \right )^2\left[1-\left(\frac{m_{\varphi}}{ m^{ (g) }_{\rm KK}}\right)^2\right]^3 \frac{ m^{ (g) }_{\rm KK}}{24 \pi} \,, \\
\Gamma ( g_{KK} \to qq ) & \approx &  \left(\frac{g^2_{g}}{ g_{g_{\rm KK}}} \right )^2\frac{m^{ (g) }_{\rm KK}}{24 \pi}\,, \label{eq:brtoqq}
\eea
where we denote the KK gluon mass by $m_{ \rm KK }^{ (g) }$ which is taken to be slightly smaller than KK graviton mass $m^{ (G) }_{\rm KK}$. 
The partial decay width to the top quark (and also the bottom quark) might depend on the contribution from the non-universal coupling piece~\cite{Agashe:2016rle,Agashe:2016kfr}, but we assume that the Higgs brane scale is much larger than the IR brane scale 
(as mentioned also in footnote~\ref{ft:footnote})
so that the partial decay width to $t\bar{t}$ (and also $b\bar{b}$) takes the form of eq.~\eqref{eq:brtoqq} up to the $m_t$-dependent phase-space factor. 

\paragraph*{($c$) KK Graviton:} Couplings of KK graviton involved in its production and decay are encoded in the following interaction terms:
\bea
\mathcal L^{\rm graviton}_{ \rm extended \; warped } & \supset & 
\frac{ g_{\rm grav} }{ m^{ (G) }_{ \rm KK } } G^{ \mu \nu }_{ \rm KK } \left( 
a T_{ \mu \nu }^{ ( \rm radion ) } 
+ b \frac{ g^2_A }{ g^2_{ A_{\rm KK } } }  T_{ \mu \nu }^{ ( \rm SM \; gauge  ) } + c  \frac{ g_A }{ g_{ A_{\rm KK} } } 
T_{ \mu \nu }^{ ( \rm SM-KK gauge ) } \right), \nonumber \\
\eea
where $a$, $b$, and $c$ are $O(1)$ coefficients and where $T_{ \mu \nu }$ is the energy-momentum tensor for the non-gravitational fields with $T_{ \mu \nu }^{ ( \rm SM-KK gauge ) }$ denoting a ``mixed'' one (i.e., for SM and KK gauge bosons).
As outlined in Appendix E of \cite{Contino:2006nn}, these couplings can be obtained as follows. 
In the composite sector and mixing terms (but {\em not} in the elementary sector terms), we include the KK graviton field as fluctuations around the metric, i.e., we simply promote $\eta_{ \mu \nu }$ to $\left[ \eta_{ \mu \nu } + \left( \dfrac{ g_{\rm grav} }{ m^{ (G) }_{ \rm KK } } G^{ \mu \nu }_{ \rm KK } \right) \right]$. 
Then, in order to get the couplings of KK graviton to the SM and KK gauge bosons, we diagonalize the spin-1 mass matrix.

The partial decay widths of KK graviton relevant to our signal channels introduced in eqs.~\eqref{eq:SR1} and \eqref{eq:SRII}, respectively, are the ones for a radion pair and for a KK gluon plus a SM gluon.
For estimating the bound on the mass of KK graviton from current searches, the decay width to a gluon pair is also important.
The analytic expressions for those three decay channels are given by 
\begin{eqnarray}
    \Gamma(G_{\rm KK}\to \varphi \varphi) &\approx&  g_{\rm grav}^2\frac{m_{\rm KK}^{(G)}}{960\pi}\left\{1-4\left(\frac{ m_\varphi}{m_{\rm KK}^{(G)}}\right)^2 \right\}^{5/2}, \label{eq:brto1}\\
    \Gamma(G_{\rm KK}\to g_{\rm KK}g) &\approx&  \left(\frac{g_g g_{\rm grav}}{g_{g_{\rm KK}}}\right)^2 \frac{m_{\rm KK}^{(G)}}{30\pi}\left\{1-\left( \frac{m_{\rm KK}^{(g)}}{m_{\rm KK}^{(G)}}\right)^2 \right\}^3\left\{6+3\left( \frac{m_{\rm KK}^{(g)}}{m_{\rm KK}^{(G)}}\right)^2+\left( \frac{m_{\rm KK}^{(g)}}{m_{\rm KK}^{(G)}}\right)^4 \right\}, \nonumber \\
    \label{eq:brto2} \\
    \Gamma(G_{\rm KK}\to gg) &\approx& \left( g_{\rm grav}\frac{g_{g}^2}{ g_{g_{\rm KK}}^2} \right )^2 \frac{m_{\rm KK}^{(G)}}{10\pi}\,, \label{eq:brto3}
\end{eqnarray}
where coefficients $a$, $b$, and $c$ are set to be unity for simplicity. 

\section{LHC Signals from KK Graviton}
\label{sec:overview}

Based on the model outlined above, in this section, we discuss the LHC signals from KK graviton of our interest and associated issues such as backgrounds and existing experimental bounds relevant to our model. 
At the end of this section, we define a set of benchmark parameter values based on which our example analyses will be performed in section~\ref{sec:analysis}.

\subsection{Experimental signatures and backgrounds}
As elaborated in the introductory section, there arise four gluons in the final state in both SR I and SR II, and hence the experimental signature is simply given by four gluon-initiated jets. 
However, the underlying event topologies are different so that different search strategies are motivated. 
The two signal regions are essentially determined by decay branching fractions of the KK graviton. So, it is instructive to compare the branching fractions into a radion pair and into a KK-gluon-SM-gluon pair. 

\begin{figure}[t]
    \centering
    \includegraphics[width=7.5cm]{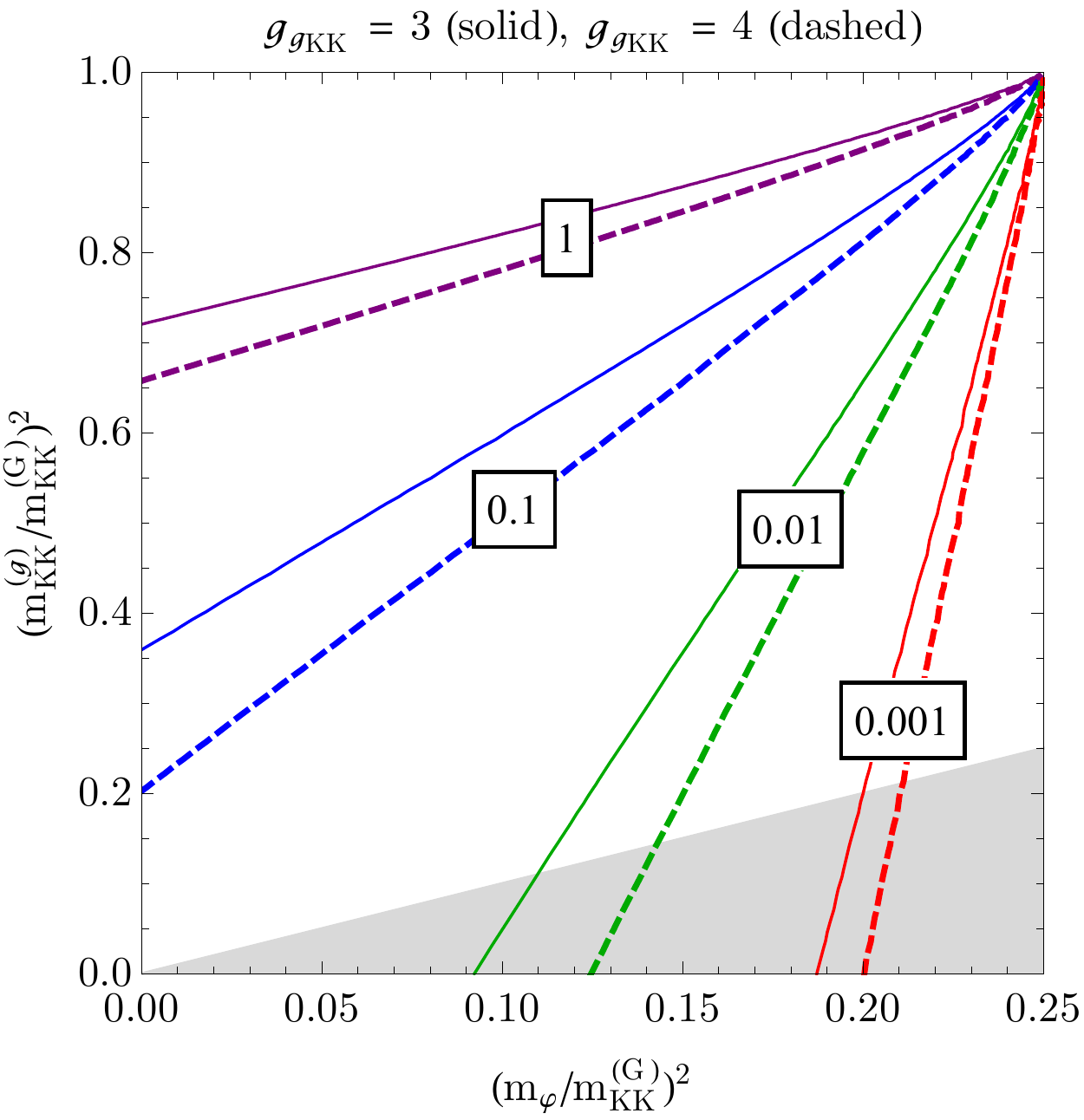}
    \includegraphics[width=7.3cm]{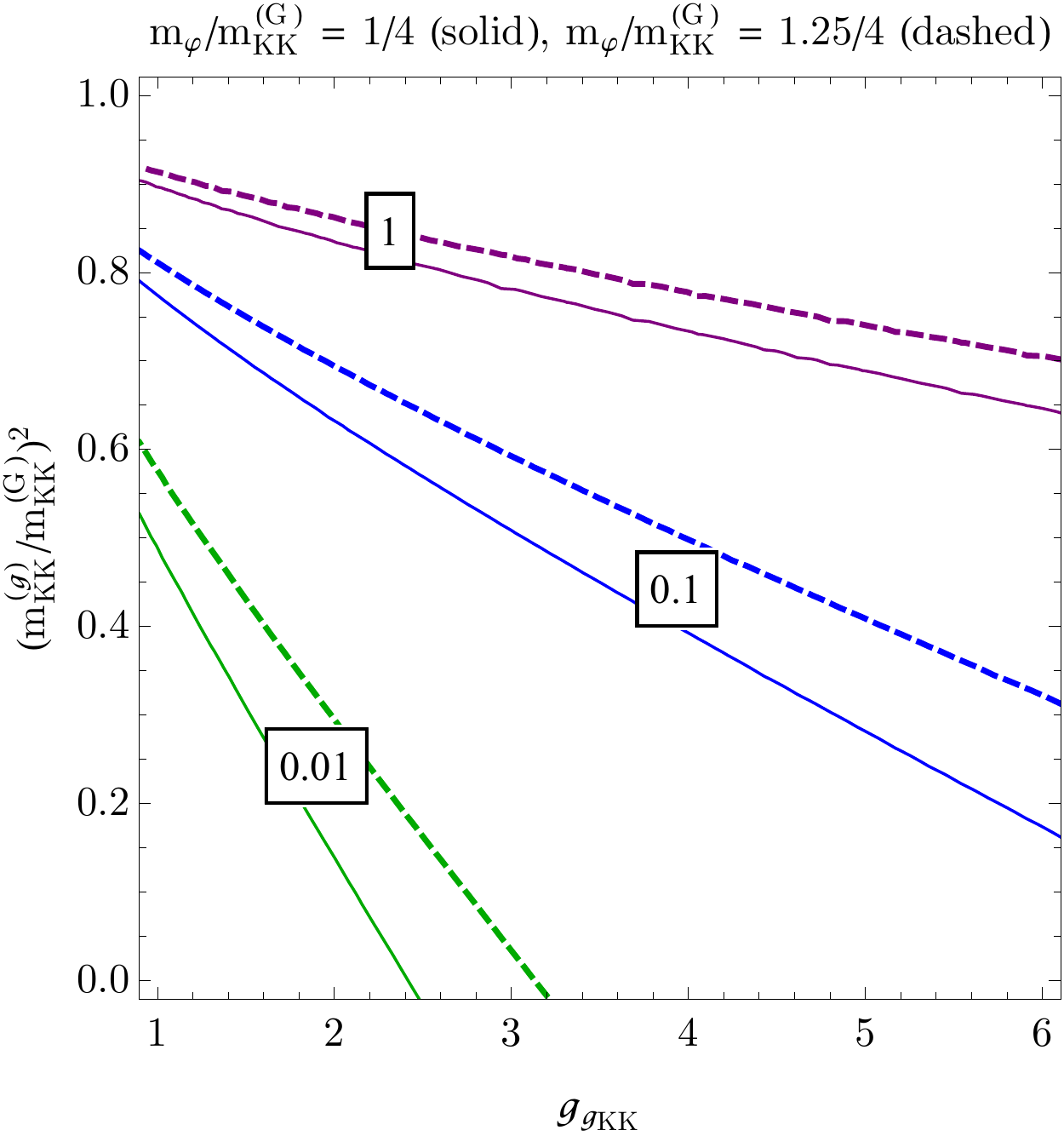}
    \caption{Left: The BRs of KK graviton to two radions relavitve to that to a KK gluon and a SM gluon in the $\left(\left[m_\varphi/m_{\rm KK}^{(G)}\right]^2,\left[m_{\rm KK}^{(g)}/m_{\rm KK}^{(G)}\right]^2\right)$ plane, i.e., each contours is defined by a constant ratios of $\Gamma(G_{\rm KK}\to \varphi \varphi)$ to $\Gamma(G_{\rm KK}\to g_{\rm KK}g)$. Coupling $g_{g_{\rm KK}}$ is fixed to two representative values, 3 (solid lines) and 4 (dashed lines). In the gray-shaded region, the decay of KK gluon to an on-shell radion and a SM gluon is not kinematically allowed. 
    Right: The corresponding plots in the $\left(g_{g_{\rm KK}},\left[m_{\rm KK}^{(g)}/m_{\rm KK}^{(G)}\right]^2\right)$ plane, while $m_\varphi/m_{\rm KK}^{(G)}$ is fixed to 1/4 (solid lines) and 1.25/4 (dashed lines).  }
    \label{fig:Br}
\end{figure}

Figure~\ref{fig:Br} displays the BRs of KK graviton to two radions relative to the BRs to a KK gluon and a SM gluon in the $\left(\left[m_\varphi/m_{\rm KK}^{(G)}\right]^2,\left[m_{\rm KK}^{(g)}/m_{\rm KK}^{(G)}\right]^2\right)$ plane (left panel) and in the $\left(g_{g_{\rm KK}},\left[m_{\rm KK}^{(g)}/m_{\rm KK}^{(G)}\right]^2\right)$ plane (right panel). The contours are defined by the ratio of $\Gamma(G_{\rm KK}\to \varphi \varphi)$ to $\Gamma(G_{\rm KK}\to g_{\rm KK}g)$ given in eqs.~\eqref{eq:brto1} and \eqref{eq:brto2}, respectively. 
In the left panel, coupling $g_{g_{\rm KK}}$ is set to two representative values, 3 (solid lines) and 4 (dashed lines), while in the right panel, $m_\varphi/m_{\rm KK}^{(G)}$ is fixed to 1/4 (solid lines) and 1.25/4 (dashed lines). 
In the gray-shaded region of the left panel, the decay of KK gluon to an on-shell radion and a SM gluon is not kinematically allowed.
The purple lines essentially play a role of boundaries to separate SR I and SR II. Above (below) the lines the radion (KK gluon) scenario comes with more parton-level production cross section than the KK gluon (radion) scenario. Also considering the existing bounds on the KK gluon and radion from dijet resonance searches that we will discuss in the next section, we see that the upper-right region of parameter space in the left panel (i.e., heavier KK gluon and radion) and the upper-right region of parameter space in the right panel (i.e., heavier KK gluon and larger $g_{g_{\rm KK}}$) are better motivated for a fixed set of the other parameters. 

Given the experimental signature, we expect that the dominant background source of our KK graviton signals is four QCD jet events.
To estimate the backgrounds we simulate events with four jets in the final state produced by the SM model processes, following the simulation scheme that we describe in section~\ref{sec:simulation}. 
Other potential background sources are SM 3-jet final states that produce four jets after showering/hadronization and the four highest $p_T$ jets in the SM 5-jet final state.
A more precise estimate would be made with an appropriate jet-matching scheme, but we do not pursue this direction in our current study, reserving it for future work. 
Our final analysis cuts are hard enough, so we expect that the contribution from 3-jet events is subleading. Likewise, we expect that the contribution from 5-jet events is subdominant. Therefore, our background estimate can be a good approximation in studying signal sensitivities, with an at most $\mathcal{O}(1)$ potential uncertainty.

\subsection{Current bounds from standard dijet searches}

We remind that all the three new particles, KK graviton, KK gluon, and radion, involved in our signal processes possess a decay channel into a pair of jets. 
Both radion and KK graviton produce a gluon-induced dijet final state while the KK gluon produces a quark-induced one. 
In regards of LHC bounds on these cross-sections (including {\em direct} radion production from gluon fusion), the CMS search in ref.~\cite{Sirunyan:2019vgj} and the ATLAS search in ref.~\cite{Aad:2019hjw} with an integrated luminosity of $\sim 140$~fb$^{-1}$ are relevent for the high-mass ($\gtrsim 1.5$~TeV) resonances, whereas the CMS search in ref.~\cite{Sirunyan:2018xlo} 
with an integrated luminosity of $\sim 40$~fb$^{-1}$ is sensitive to the low-mass ($\lesssim 1.5$~TeV) resonances. 

\begin{figure}[t]
    \centering
    \includegraphics[width=7.5cm]{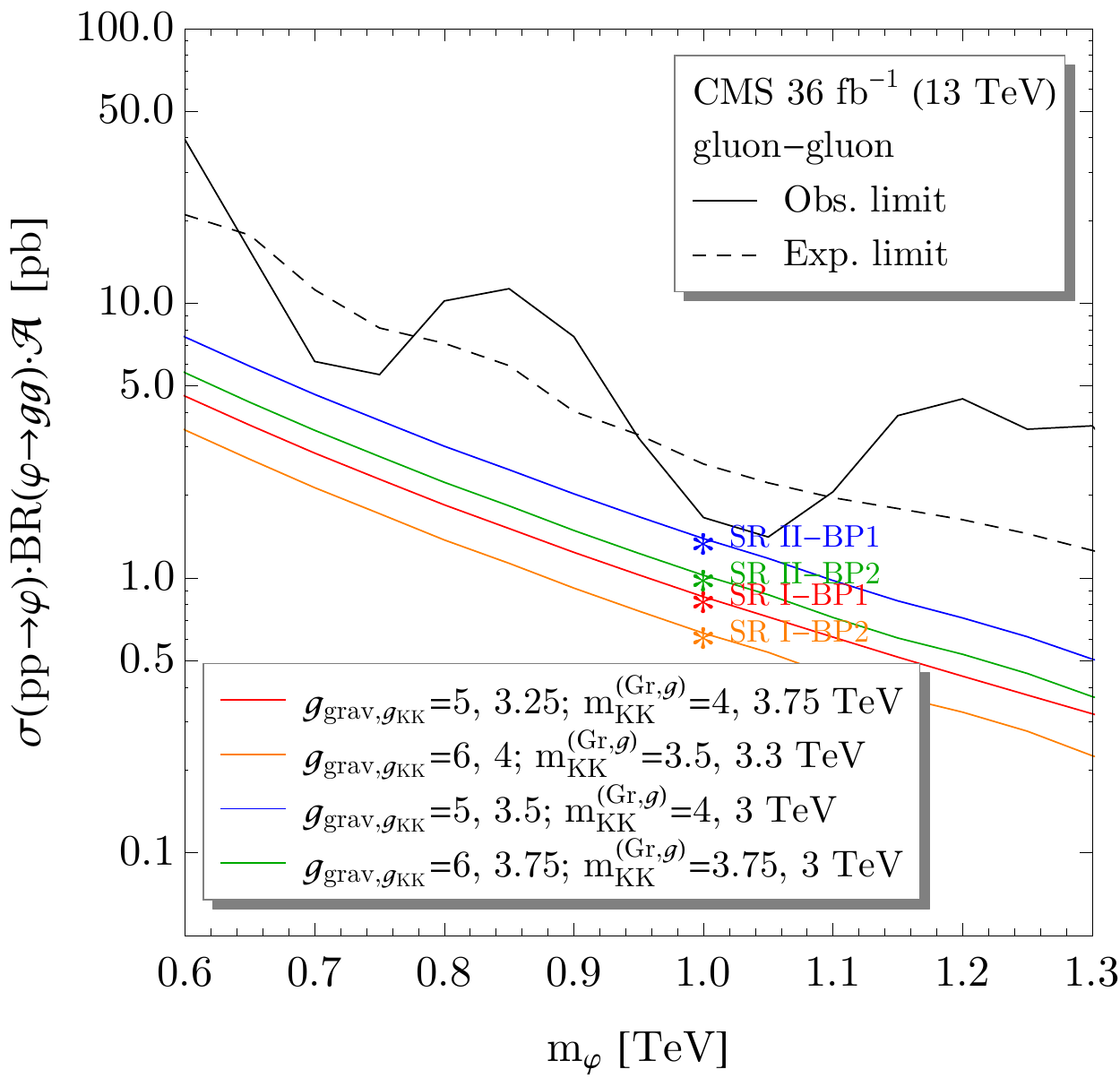}
    \includegraphics[width=7.5cm]{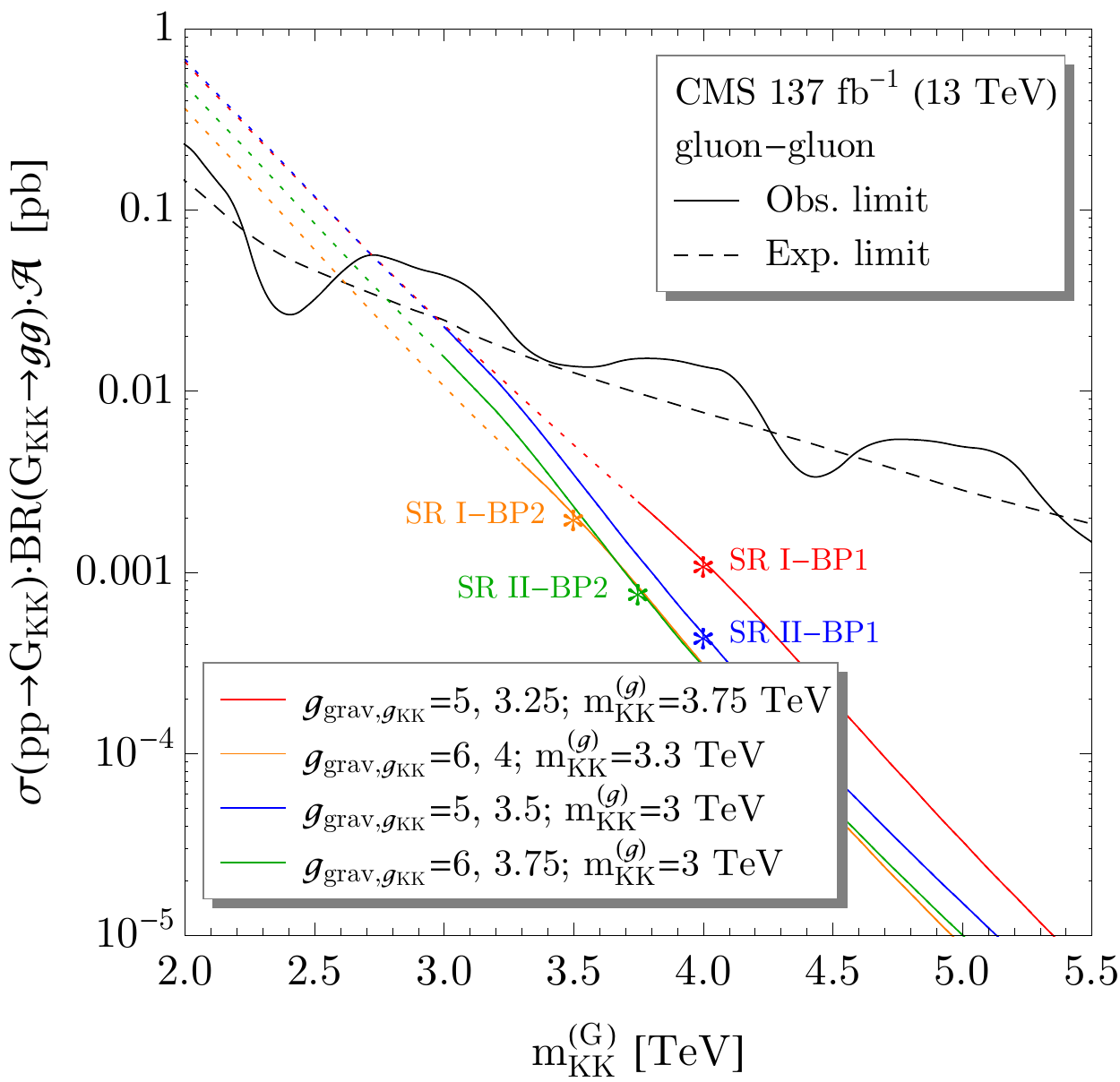} \\ 
    \includegraphics[width=7.5cm]{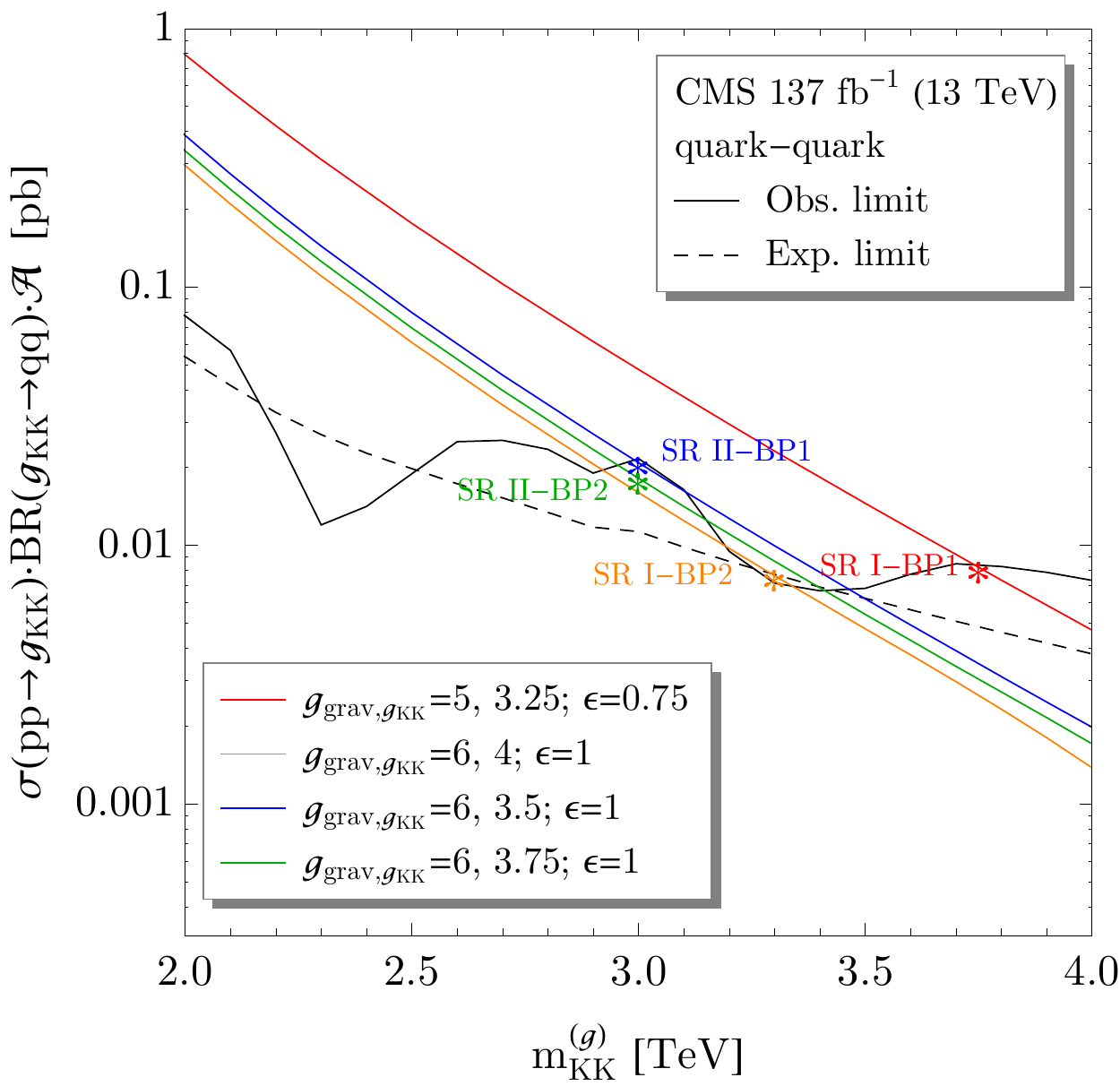}
    \caption{Top-left: $gg$-produced low-mass ($\lesssim 1.5$~TeV) resonance bounds from the CMS dijet search with an integrated luminosity of 36~fb$^{-1}$ at $\sqrt{s}=13$~TeV~\cite{Sirunyan:2018xlo}. The color-coded lines are the fiducial cross-sections of radion production for four sets of parameter choices on which our benchmark points in table~\ref{tab:BP} are marked by asterisk symbols. The BR of radion to two SM gluons is (almost) 100\%, while an acceptance of 100\% is applied for conservative estimates.   
    Top-right: $gg$-produced high-mass ($\gtrsim 1.5$~TeV) resonance bounds from the CMS dijet search with an integrated luminosity of 137~fb$^{-1}$ at $\sqrt{s}=13$~TeV~\cite{Sirunyan:2019vgj}. The color-coded lines are the fiducial cross-sections of KK graviton production with $m_\varphi$ set to be 1 TeV for four sets of parameter choices on which our benchmark points are marked by asterisk symbols. The segments drawn by dotted lines are for the situation of $m_{\rm KK}^{(G)}<m_{\rm KK}^{(g)}$ that is theoretically disfavored. 
    Bottom: $qq$-produced high-mass ($\gtrsim 1.5$~TeV) resonance bounds from the CMS dijet search with an integrated luminosity of 137~fb$^{-1}$ at $\sqrt{s}=13$~TeV~\cite{Sirunyan:2019vgj}. The color-coded lines are the fiducial cross-sections of KK gluon production for four sets of parameter choices with $m_\varphi=1$~TeV on which our benchmark points are marked by asterisk symbols.}
    \label{fig:limits}
\end{figure}

A produced radion decays to a pair of gluon-induced jets with a BR of $\sim 100\%$. Since we are interested in relatively light radion, a CMS resonance search in the dijet channel at $\sqrt{s}=13$~TeV~\cite{Sirunyan:2018xlo} can constrain the radion mass parameter.
The top-left panel of figure~\ref{fig:limits} exhibits the reported CMS bounds (black lines) in $gg$-produced resonance mass and expected parton-level radion production cross-sections multiplied by the BR and acceptance $\mathcal{A}$ (color-coded lines). Here the production cross sections and the BRs are evaluated by \texttt{MG5@aMC}~\cite{Alwall:2014hca} (see section~\ref{sec:simulation} for more detailed information on our simulation scheme), while the acceptance is conservatively assumed to be 100\% for all mass points. 
To develop the intuition, we show radion production cross-sections for several parameter choices described in the legend. We also mark our benchmark points, which we will define in the next subsection, by the asterisk symbols which are placed below the current limits. 
Cross-sections for other choices of coupling constants can be easily obtained by appropriately rescaling the ones that we report here.  

The KK graviton is created dominantly by the gluon fusion and can decay into the gluon-initiated dijet final state. 
The BR calculation involves all of the model parameters except $\epsilon$ [see eqs.~\eqref{eq:brto1} through \eqref{eq:brto3}], but for the parameter regions of our interest, i.e., $m_{\rm KK}^{(G)} \gtrsim m_{\rm KK}^{(g)} > m_\varphi$, the BR to a gluon pair is $\sim50-60\%$. Since we are interested in relatively heavy KK graviton, high-mass resonance searches in the dijet channel~\cite{Sirunyan:2019vgj,Aad:2019hjw} can provide the limits on the KK graviton mass parameter. 
The top-right panel of figure~\ref{fig:limits} displays the reported CMS bounds\footnote{The corresponding ATLAS search~\cite{Aad:2019hjw} also provides similar bounds. However, ATLAS assumed the resonance shape to be approximated to the Gaussian, whereas CMS took a skewed shape, considering the shape distortion by detector effects in the jetty final state.} (black lines) again in the $gg$-produced resonance mass and expected parton-level KK graviton production cross-sections multiplied by acceptance, which is again assumed to be 100\%, and the associated BR (color-coded lines). For all colored lines, the mass of radion is set to be 1 TeV, and since $m_{\rm KK}^{(G)}<m_{\rm KK}^{(g)}$ is theoretically disfavored, the corresponding segments are drawn by dotted lines. As before, calculation of cross-sections and BRs is done with \texttt{MG5@aMC}, and a few lines of KK graviton production cross-sections are presented for developing the intuition together with our benchmark points marked by the asterisk symbols and placed below the current limits. 

Finally, when it comes to the limits for KK gluon, we remark that they are qualitatively different from those for KK graviton and radion. First, KK gluon is mostly produced via $q\bar{q}$ annihilation, and the dijet final state is induced by quarks. In principle, the final states of heavy quark flavors $b$ and $t$ could be distinguished from those of light quark flavors by the $b$ tagging and the additional decay signature of $t$, respectively. Note that top quark decay products would be highly collimated, hence appear as a single (boosted) jet due to mass gap between KK gluon and top quark. In our estimate we consider neither $b$ tagging nor $t$-jet tagging, i.e., the di-light-quark and the di-bottom/top final states are treated on an equal footing, so the limits for KK gluon are rather conservative.  
Second, the BR to a quark pair is determined by all three coupling constants, $g_{\rm grav}$, $g_{g_{\rm KK}}$, and $\epsilon$ together with $m_{\rm KK}^{(g)}$ and $m_\varphi$, and it varies quite sensitively according to coupling choices. Therefore, more care should be taken to check whether or not a given $m_{\rm KK}^{(g)}$ is safe from the existing bounds. 
We show the reported CMS bounds (black lines) in the $qq$-produced resonance mass and expected parton-level KK gluon production cross-sections multiplied by acceptance and BR (color-coded lines) in the bottom panel of figure~\ref{fig:limits}. Again, several lines of KK gluon production cross-sections are shown for developing the intuition, with our benchmark points marked by the asterisk symbols, while $m_\varphi$ is fixed to be 1 TeV. Note that the four colored lines are valid as far as KK graviton is kept heavier than KK gluon as the decay channel to KK graviton would open otherwise.\footnote{Of course, $m_{\rm KK}^{(G)}>m_{\rm KK}^{(g)}$ is theoretically preferred.} We see that all of our benchmark points are below the CMS observed limits. 

\subsection{Benchmark points}

We are now in the position to define our benchmark points (BPs) that are investigated in section~\ref{sec:analysis}, predicated upon the considerations in the previous subsections. The parameter choices for our BPs are summarized in table~\ref{tab:BP}, with the mass values given in TeV. As discussed previously, the antler decay and the sequential cascade decay are dominant in SR I and SR II, respectively, so we associate each SR with the dominating scenario in the parentheses. All the benchmark points listed in the table are consistent with the existing LHC bounds. We also report the expected statistical significance, which is defined by the number of signal events divided by the square-root of the number of background events, at the LHC14 with a luminosity of 3,000~fb$^{-1}$ in the last column as a preview, while we shall detail our analysis strategy and results in sections~\ref{sec:simulation} and \ref{sec:analysis}. 

\begin{table}[t]
    \centering
    \resizebox{\columnwidth}{!}{%
    \begin{tabular}{c|c|c c c c c c | c}
    \hline \hline
        \multicolumn{2}{c|}{} & $g_{\rm grav}$ & $g_{g_{\rm KK}}$ & $\epsilon$ &  $m_{\rm KK}^{(G)}$ & $m_{\rm KK}^{(g)}$ & $m_\varphi$  & Significance\\
    \specialrule{.1em}{.05em}{.05em}
    \multicolumn{1}{ c|  }{\multirow{2}{*}{SR I (antler)} } &
    \multicolumn{1}{ c| }{BP1} & 5 & 3.25 & 0.75 & 4 & 3.75 & 1 & 4.1 (table~\ref{Tab:Antler_BP1})\\
    \multicolumn{1}{ c|  }{}                        &
    \multicolumn{1}{ c| }{BP2} & 6 & 4 & 1 & 3.5 & 3.3 & 1 & 4.7 (table~\ref{Tab:Antler_BP2})\\
    \hline
    \multicolumn{1}{ c|  }{\multirow{2}{*}{SR II (sequential cascade)} } &
    \multicolumn{1}{ c| }{BP1} & 6 & 3.5 & 1 & 4 & 3 & 1  & 2.5 (table~\ref{Tab:Cascade_BP1})\\ 
    \multicolumn{1}{ c|  }{}                        &
    \multicolumn{1}{ c| }{BP2} & 6 & 3.75 & 1 & 3.75 & 3 & 1 & 2.9 (table~\ref{Tab:Cascade_BP2}) \\
    \hline \hline
    \end{tabular}
    }
    \caption{A summary table of our benchmark points that are investigated in section~\ref{sec:analysis}, with the mass parameters given in TeV. 
    The antler decay and the sequential cascade decay are dominant in SR I and SR II, respectively. 
    All the benchmark points listed here are safe from the existing bounds (see also figure~\ref{fig:limits}).
    In the last column, we preview the expected statistical significance, which is defined by the number of signal events divided by the square-root of the number of background events, at the LHC 14 TeV with a luminosity of 3,000~fb$^{-1}$ (see section~\ref{sec:analysis} for more details).  
}
    \label{tab:BP}
\end{table}

For SR I, we choose 4~TeV and 3.5~TeV KK gravitons with $g_{\rm grav}=5$ and $g_{\rm grav}=6$, respectively. Large values of $g_{\rm grav}$ allow us to get enough signal statistics, while smaller, hence more perturbative, $g_{\rm grav}$ might suffice for mass choices of KK graviton smaller than our BPs.
KK gluon is taken to be slightly lighter\footnote{such a choice can satisfy -- even if barely -- the bound on KK gluon from standard dijet search which might be $\sim 3$ TeV.} so that the decay width of KK graviton to a KK gluon and a SM gluon is suppressed by a lack of phase space although the associated coupling is sizable (i.e., intermediate to those to two SM gluons and two radions), thus rendering the associated decay BR small. 
The radion mass parameter is chosen to be above the bound from dijet search (based on its direct production) so as to minimize the phase-space suppression of this decay mode. 
For SR II, we choose $m_{\rm KK}^{(G)}$ to be 4 and 3.75 TeV with $g_{\rm grav}=6$. 
However, $m_{\rm KK}^{(g)}$ is smaller than $m_{\rm KK}^{(G)}$ by $\sim 20-25\%$ so that the phase space for the decay to a KK gluon opens up sufficiently. 
As a result, this decay mode can dominate with its coupling being larger than the other kinematically opened decay channel, i.e., into two SM gluons.

It is noteworthy that while the values of the couplings $g_{ g_{ \rm KK } }$ and $g_{ \rm grav }$ chosen above are larger than those encountered typically in the SM, they can nonetheless be ``perturbative'' in the sense that the corresponding loop expansion parameter, $\sim\dfrac{ g_{  g_{ \rm KK }, \; \rm grav }^2}{16 \pi^2}$ 
is still less than 1.
Indeed, such sizable couplings for KK modes are perhaps ``expected'' based on their interpretation as composites of a new strong sector
(via the AdS/CFT duality): in this context recall that the coupling of $\rho$ meson to $\pi$'s in real QCD/hadronic sector is $g_{ \rho }\sim 6$ (i.e., similar values to what we have taken for the extended warped model), with the associated hadronic loop expansion parameter being $\sim \dfrac{g_{ \rho }^2}{16 \pi^2 }\sim \dfrac{1}{N_c}$, i.e., large $N_c$ expansion for hadrons is often used
here, even though $N_c = 3$ (as is well-known such a $1 / N$ expansion is dual to 5D/KK loops).

\section{Event Simulation and Selection \label{sec:simulation}}

Before presenting the analysis of the above benchmark points, in this section, we detail our event simulation scheme and discuss general 
selection criteria for isolating signal events from background events. 
We perform Monte Carlo simulation of proton collisions at the LHC to predict the effects of the KK graviton channels and to estimate the background from SM 4-jet final state processes. Signals and background processes are simulated at $\sqrt{s}=14$~TeV.  The full chain of the simulation takes into account parton showering, hadronization, and detector-level effects. The simulation begins by generating parton events with \texttt{MG5@aMC}~\cite{Alwall:2014hca} and using parton distribution function (PDF) set \texttt{NNPDF2.3} QCD NLO~\cite{Ball:2012cx}. 
The output from \texttt{MG5@aMC} is pipelined to \texttt{Pythia6.4}~\cite{Sjostrand:2006za} for showering and hadronization and \texttt{Delphes3}~\cite{deFavereau:2013fsa} for detector responses. 
Jets are reconstructed by \texttt{FastJet}~\cite{Cacciari:2011ma} implemented in \texttt{Delphes3}, based on the anti-$k_t$ algorithm with jet radius parameter $R$ being 0.5.
The output from \texttt{Delphes3} is fed into our analysis, where cuts on kinematic variables are imposed on signal and background events. 

When generating signal events, we include both antler and sequential-cascade topologies in both signal regions, although one may significantly dominate over the other. 
We find that depending on the mass spectrum choices, a set of cuts optimized to the event topology whose production cross section is the larger can result in a better acceptance for the smaller, resulting in a comparable contribution to the total number of signal events. We will revisit this interesting possibility in the next section together with a concrete benchmark point. 
For the 4-jet QCD background, we generate events with a set of ``pre-selection'' cuts at the parton level, for a practical purpose.
They are sufficiently softer than our analysis cuts, so they effectively allow us to have enough background statistics within relevant phase space. 
We will discuss our choices in the next section in the context of our benchmark points. 

Since a signal event in both SR I and SR II is expected to have four hard gluons in the final state, we first require it to have no isolated electromagnetic objects (i.e., no photons, no electrons, and no muons) and at least four isolated jets, i.e., $N_j \geq 4$, with transverse momentum $p_{T,j}$ being greater than 20~GeV, pseudo-rapidity $\eta_j$ lying in-between $-5$ and $5$, and no other additional activities recorded within $\Delta R=0.4$.  
However, our $p_{T,j}$- and $\eta_j$-related cuts in the analysis level are much harder than the above choices, they can be regarded as ``minimum'' requirements to pass a given signal event to more dedicated requirements. 
Moreover, since the background events are generated with ``pre-selection'' cuts mentioned above, these ``minimum'' requirements are not much relevant to them. However, again our analysis cuts are hard enough, so we expect that the detailed $p_{T,j}$ and $\eta_j$ choices 
should not affect our final results and conclusions.  

In our analysis we introduce several kinematic variables to distinguish the signal from the background. 
First of all, the signal processes in both of the signal regions involve four jets in the final state, which are produced from resonant production and decay of KK graviton, so we expect that the 4-jet invariant mass should return the mass of KK graviton: 
\beq
M^{(4j)}\equiv \sqrt{\left(\sum_{i=1}^{4} p_i\right)^2}\approx m_{\rm KK}^{(G)},
\eeq
where $p_i$ denotes the four-momentum of the $i$th jet with $i$ running over the first four hardest jets. 
Here and henceforth capital $M$ symbolizes a reconstructed mass, while lower-case $m$ is the corresponding true (or input) mass.   
As discussed previously, since in SR I two radions are produced from the decay of KK graviton [see also figure~\ref{fig:evetopo}($a$)], two (non-overlapping) pairs of jets should give rise to invariant mass values not only close to each other but close to the mass of radion. That is, 
\beq
M^{(2j)}_{ab}\equiv \sqrt{\left(p_a+p_b\right)^2} \approx M^{(2j)}_{cd}\equiv \sqrt{\left(p_c+p_d\right)^2} \approx m_\varphi\,, \label{eq:invmass2body}
\eeq
where $(ab)$ and $(cd)$ denote any pairings that satisfy the above condition.
By contrast, the signal process in SR II contains another resonance, KK gluon, on top of those of KK graviton and radion [see also figure~\ref{fig:evetopo}($b$)], so (at least) one set of three jets is expected to satisfy
\begin{equation}
    M^{(3j)}_{abc}\equiv \sqrt{\left( p_a+p_b+p_c \right)^2} \approx m_{\rm KK}^{(g)},\,\,\, M^{(2j)}_{ab}= \sqrt{\left(p_a+p_b\right)^2} \approx m_\varphi\,, \label{eq:invmass3body}
\end{equation}
where again $(abc)$ and $(ab)$ denote any groupings to meet the above requirement. 

A few comments should be made for the invariant mass variable. 
First, we remark that not only all the resonances come with sizable decay widths (especially, $\Gamma_{G_{\rm KK}}\sim 3 - 6 \%\times m_{\rm KK}^{(G)}$), but the jetty final states bring in detector smearing in reconstructed invariant masses. 
Therefore, somewhat wide invariant mass window cuts are practically imposed to capture enough signal events. 
Second, we find that the resonance peaks are shifted to lower values than the nominal ones mostly for the KK graviton. 
The reason is that final-state radiation (FSR) jets carry away some fraction of energy so that keeping four hardest jets is unlikely to return the input value.
On the other hand, some of hard enough initial-state radiation (ISR) can give rise to a large invariant mass value, developing the upper tail in the associated invariant mass distribution.
However, as we will discuss shortly, central jets (i.e., $|\eta_j|<2.5$) are selected, thus the effect of ISR is not large enough to offset the effect of FSR.
Finally, we further find that the invariant mass distributions appear mildly skewed to the lower side especially for KK graviton due to rapid rising of the gluon PDF toward the lower $\sqrt{\hat{s}}$. 
Our invariant mass cut choices are based on all these observations. 

While the invariant mass window cuts are powerful in discriminating signal events from background ones, we find that the angular distributions are useful to further separate signal events from background events. 
Especially for the events in the radion channel, the angular distribution of radions produced by the KK graviton decay has a characteristic structure.
Reference \cite{Agashe:2007zd} derived the angular distribution of the pair-produced longitudinal SM gauge bosons via an $s$-channel KK graviton exchange, using helicity amplitudes~\cite{Park:2001vk}.
A similar relation applies here for the decay to a radion pair, and we find that
\begin{equation}
    \frac{d\sigma(gg\to G_{\rm KK}\to \varphi \varphi)}{d\cos\theta^*_\varphi} \propto (1-\cos^2\theta_\varphi^*)^2\,, \label{eq:anglespec}
\end{equation}
where $\theta_\varphi^*$ defines the angle between an incoming gluon and an outgoing radion in the initial parton center-of-mass frame. 
This implies that when moving onto the center-of-mass frame, radions are likely to populate in the central region, while QCD background events lie more densely in the forward and backward regions.  
Note that this distribution relies crucially on the spin of the particles involved in the process, and particularly on the spin-2 of KK graviton. For the KK gravitons produced from gluon-gluon fusion the spin component parallel to the beam direction is $\pm 2$~\cite{Park:2001vk,Agashe:2007zd}. On the other hand, the final state particles are spinless, so their outgoing direction is preferentially beam-transverse due to the angular momentum conservation, as also supported by the above relation. 

In addition to the above-discussed invariant mass and angle variables, transverse momenta $p_{T,j}$ and pseudo-rapidity $\eta_j$ are also useful variables to enhance the signal-over-background. 
First, since signal activities are likely to be found in the central region, we choose jets with $|\eta_j|<2.5$ in both signal regions, as mentioned previously.  
Second, the mass values of KK graviton allowed by existing searches are large, so signal jets are expected to be sufficiently hard. 
Moreover, we find that our data analysis benefits from $p_T$-ordered cuts, showing detailed choices in the context of our benchmark points. 

\section{Results of LHC Signal Analyses}
\label{sec:analysis}

Having outlined the plan of our analysis, in this section, we present the results for LHC signals arising from production and subsequent decay of KK graviton. As discussed in section~\ref{sec:overview}, if the mass difference between the KK graviton and the KK gluon is small, the decay of KK graviton to a KK gluon and a SM gluon is suppressed and the 4-jet signal is dominated by the process where KK graviton decays to two radions and each radion decays to two gluons, i.e., SR I. 
See also figure~\ref{fig:Br}.
On the other hand, if the KK gluon mass is smaller, the decay of KK graviton to a KK gluon and a SM gluon is significant, and it can even be the dominant decay channel for KK graviton, i.e., SR II. 
We will show the details of our analysis and results in sections~\ref{sec:analysisantler} and \ref{sec:analysiscascade}, employing two benchmark parameter sets for each of these two signal regions. These benchmark parameter sets were introduced in table~\ref{tab:BP}, and as discussed in section~\ref{sec:overview} are consistent with the current experimental bounds. 
We will then end this section with comparing the reach for KK graviton in these 4-jets decay channels to that of standard dijet searches in section~\ref{sec:analysisdiscussion}.

\subsection{SR I: antler scenario \label{sec:analysisantler}}

\begin{figure}[t]
    \centering
    \includegraphics[width=7.3cm]{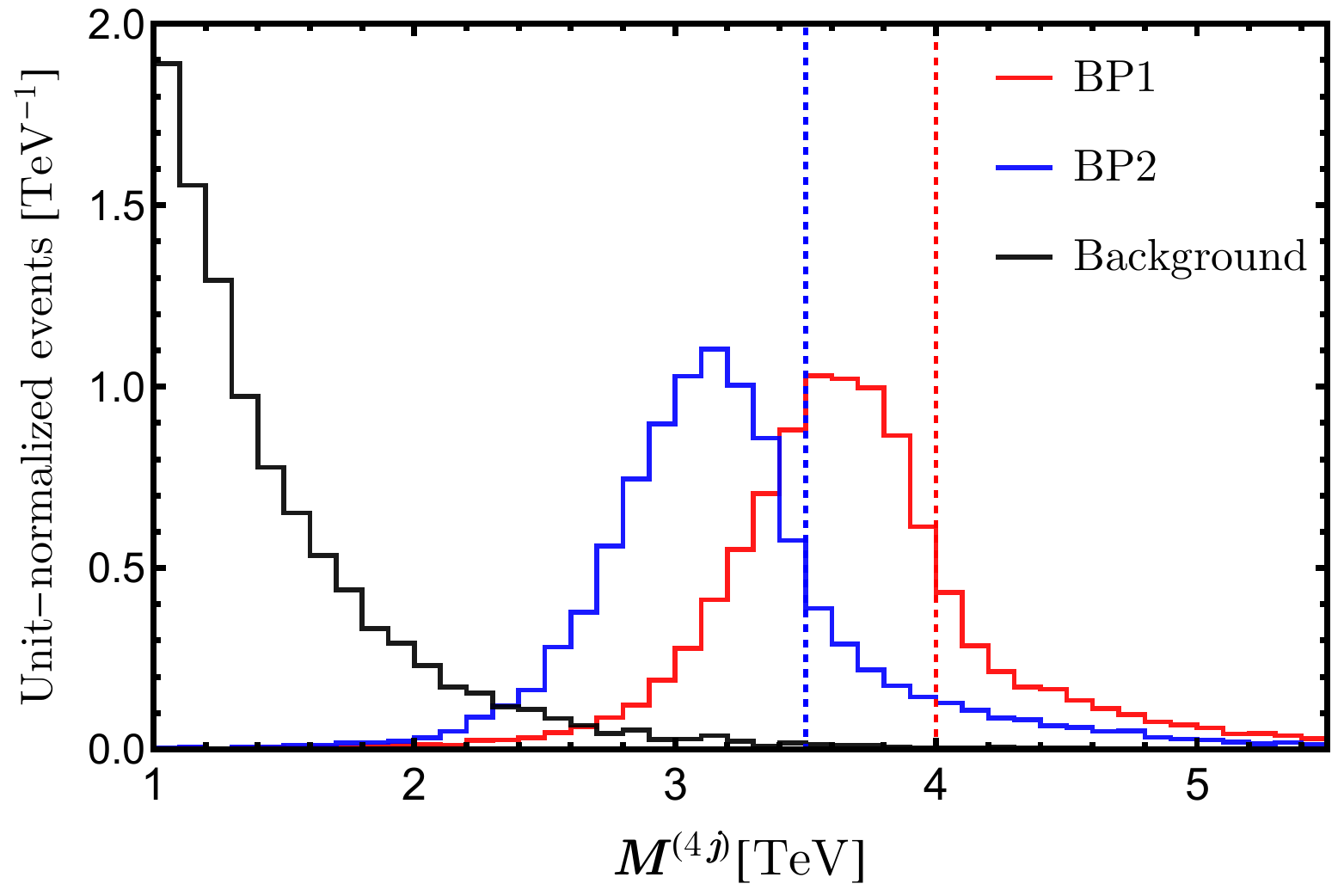} \label{fig:m4jdist_antler}
    \includegraphics[width=7.3cm]{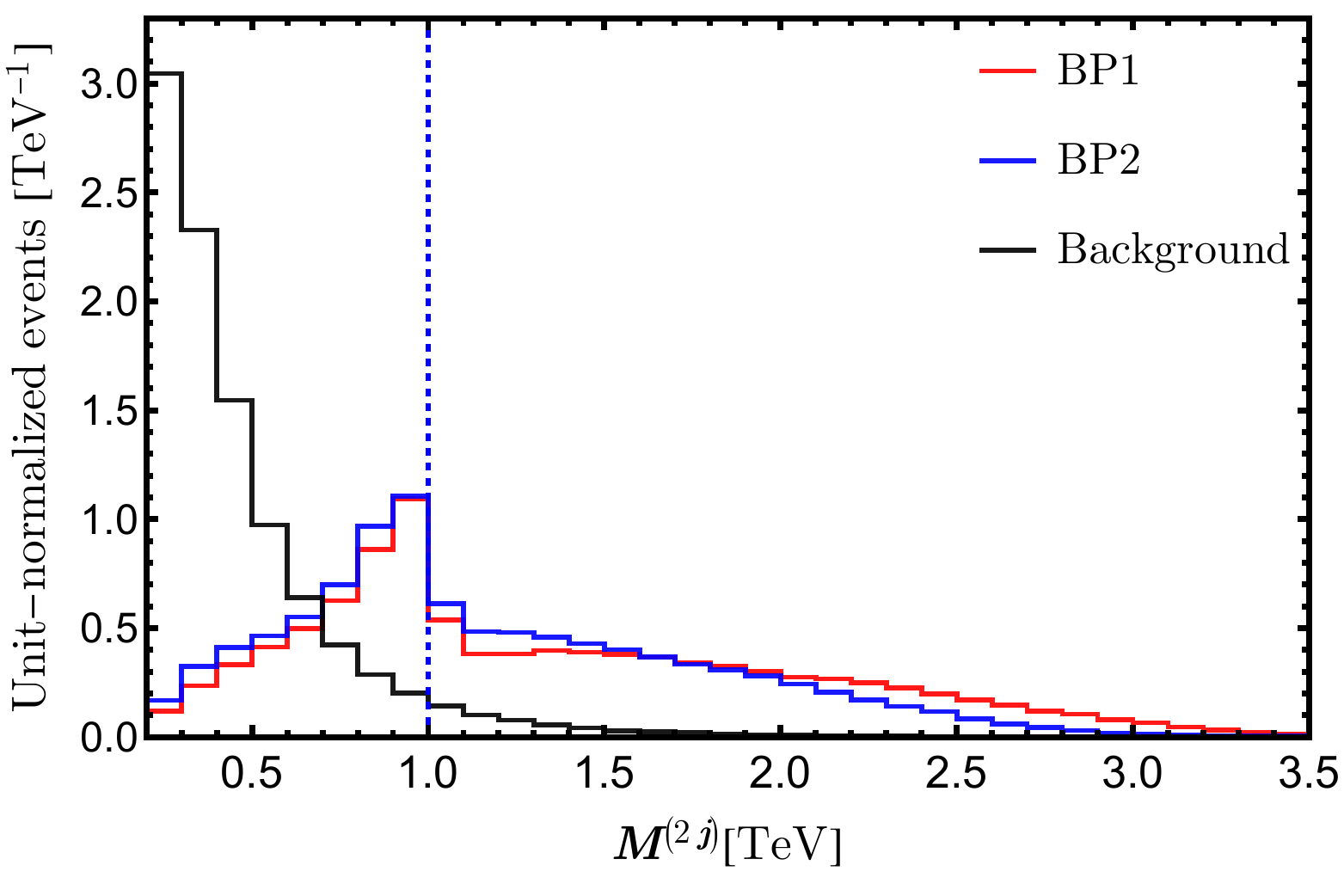} \label{fig:mjjdist_antler} \\
    \includegraphics[width=7.5cm]{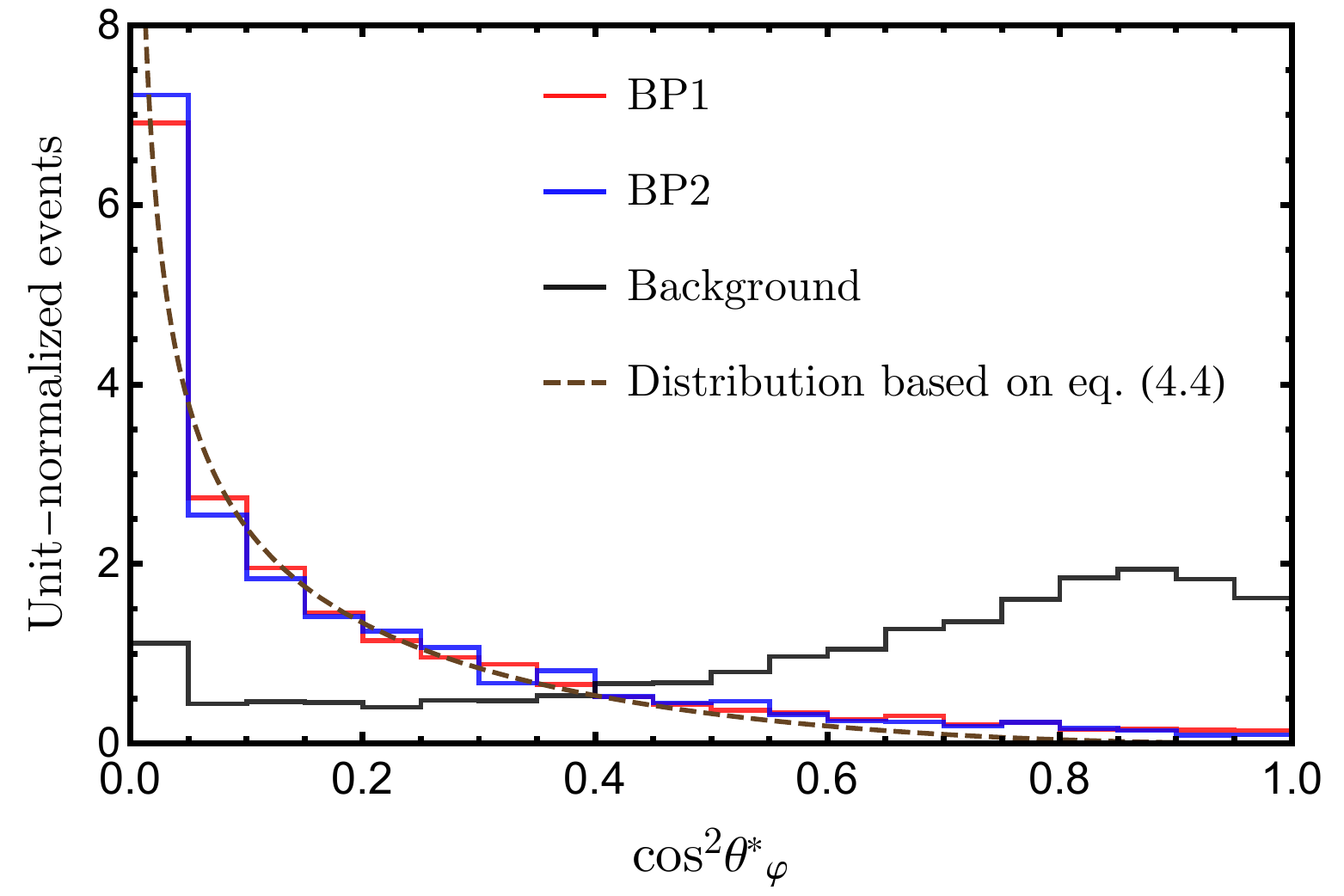} \label{fig:radangle1}    \caption{Top left: Unit-normalized $M^{(4j)}$ distributions of BP1 (red) and BP2 (blue) in SR I and the QCD 4-jet background (black). The red and blue dotted lines mark the input masses of KK graviton for BP1 and BP2, respectively.
    Top right: Unit-normalized $M^{(2j)}$ distributions of BP1 (red), BP2 (blue), and the background (black). For each event, all 6 possible dijet invariant masses have been included. 
    The dotted line marks the input mass of radion. 
    Bottom: Unit-normalized $\cos^2\theta_\varphi^*$ distributions of BP1 (red), BP2 (blue), and the background (black). The brown dashed line shows the theoretical expectation for the signal as described in eq.~\eqref{eq:anglespec}.}
    \label{fig:obsSR1}
\end{figure}

As discussed in section~\ref{sec:simulation}, the invariant mass variables are powerful in discriminating signal events from background ones. 
Since the four jets in the signal are from the decay of KK graviton, we expect that the distribution of 4-jet invariant mass $M^{(4j)}$ shows a peak near the true KK graviton mass $m_{\rm KK}^{(G)}$.  
The invariant mass is formed by the four hardest jets in their transverse momentum $p_T$. 
We show the unit-normalized $M^{(4j)}$ distributions of BP1 (red) and BP2 (blue) in SR I and the QCD background (black) in the top-left panel of figure~\ref{fig:obsSR1}. 
The dotted lines mark the input masses of KK graviton (here 4 TeV and 3.5 TeV, respectively) and we find that the signal distributions show a peak at a value slightly smaller than the input $m_{\rm KK}^{(G)}$, as discussed in the previous section.  
This simulation result motivates our choices for the 4-jet invariant mass window cuts. 

\begin{figure}[t]
    \centering
    \includegraphics[width=7.5cm]{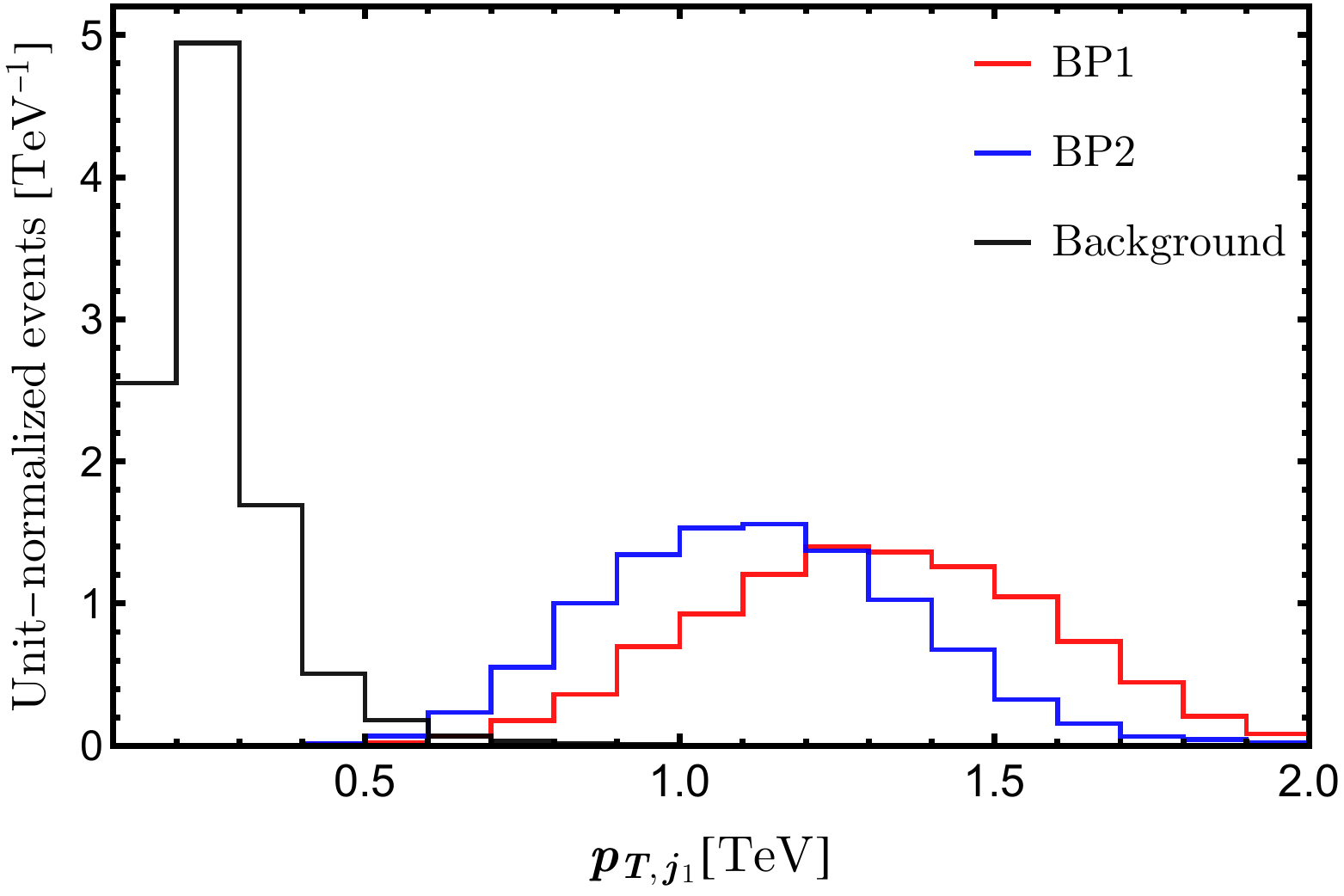} 
    \includegraphics[width=7.5cm]{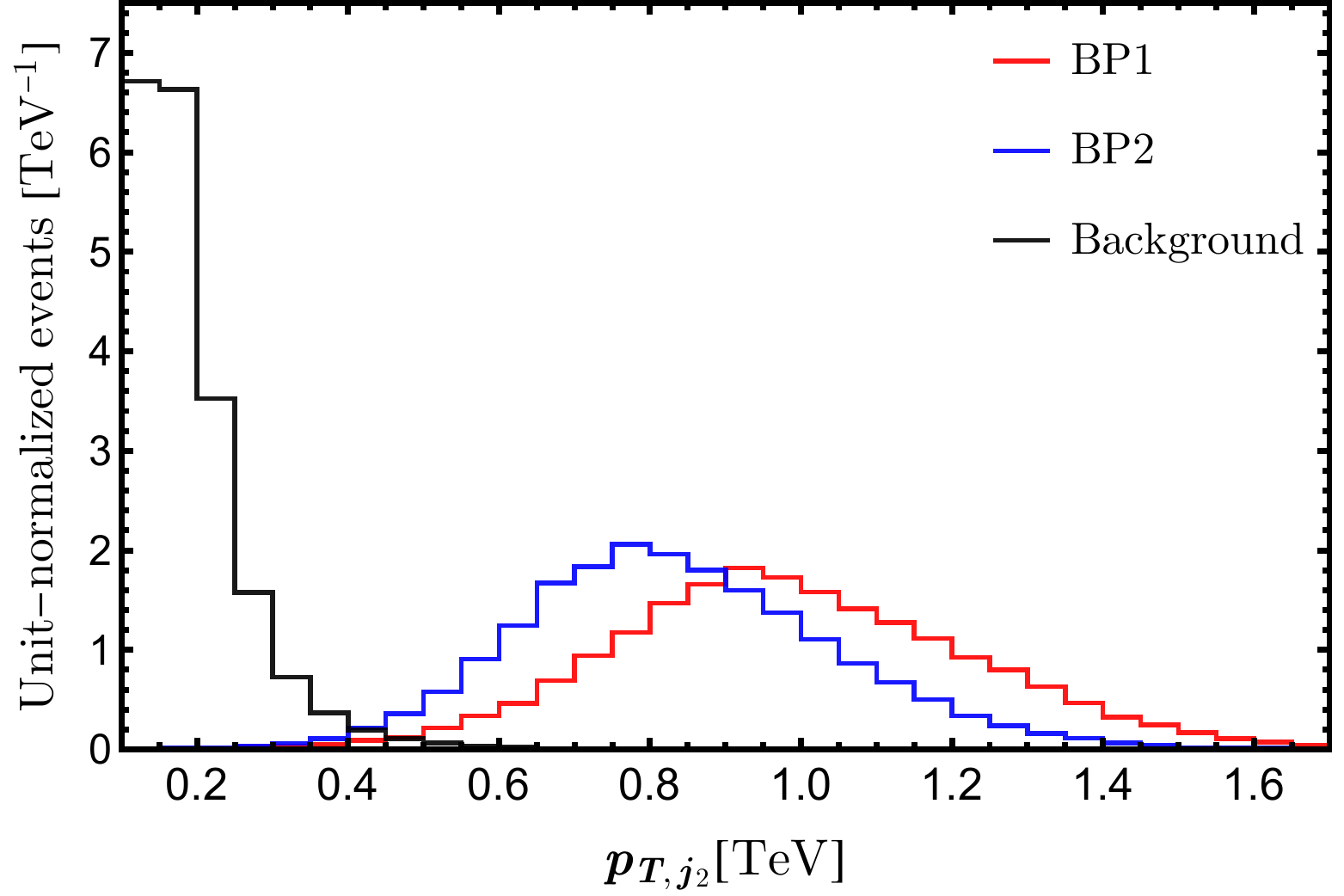} \\
    \includegraphics[width=7.5cm]{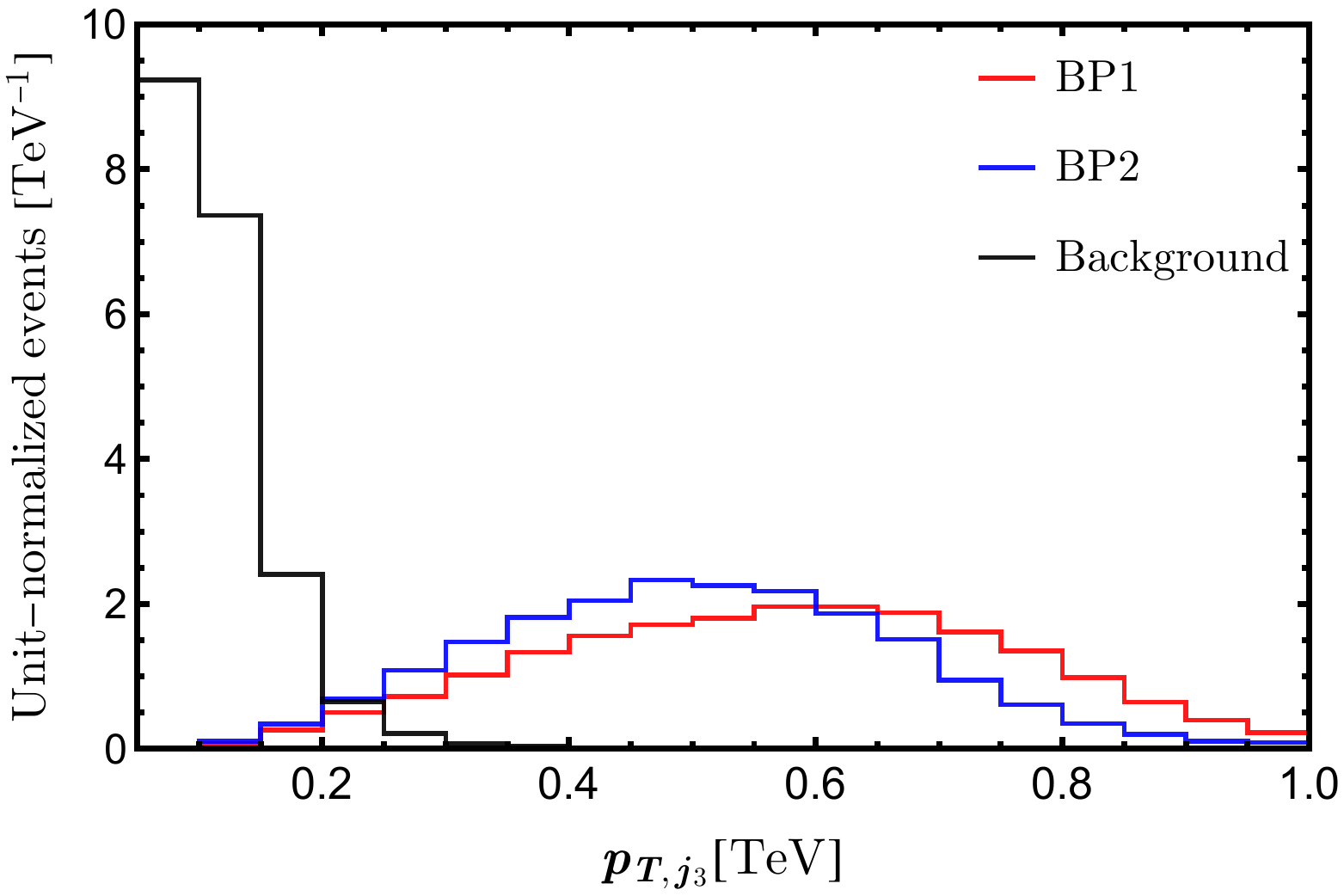} 
    \includegraphics[width=7.5cm]{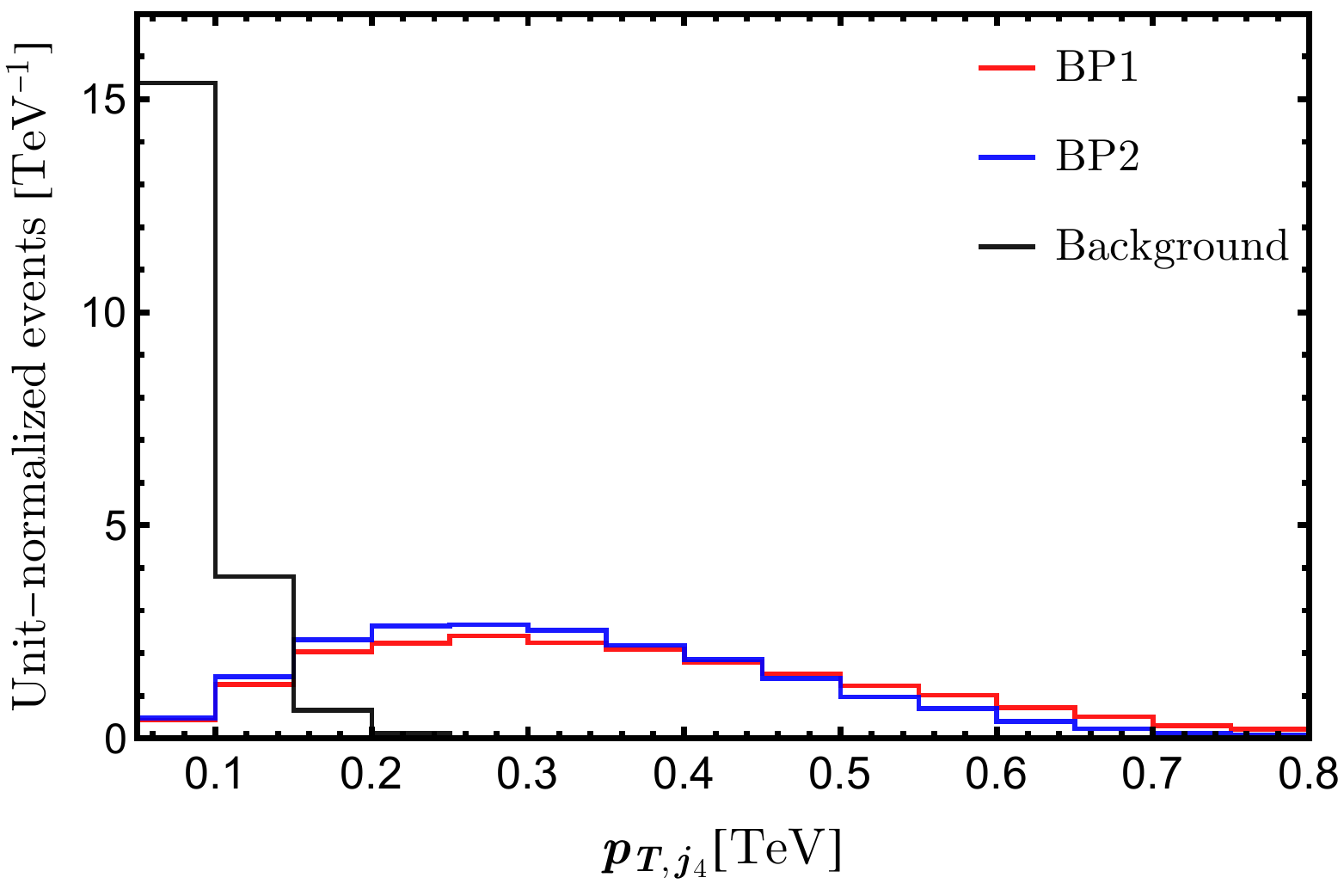}
    \caption{Distributions of ordered transverse momenta, $p_{T,j_i}$ for BP1 (red) and BP2 (blue) in SR I and the QCD 4-jet background (black).} 
    \label{fig:ptordereddist}
\end{figure}

\begin{table}[t]
\centering
\begin{tabular}[c]{ l | c | c | c }
\hline \hline
Antler cuts on BP1 in SR I & KK Gluon & Radion & QCD 4 jets \\
\specialrule{.1em}{.05em}{.05em} 
No cuts & 0.084 & 1.3 & $1.0\times10^{10}$ \\
Pre-selection cuts & 0.064 & 1.1 & 61,000 \\
$N_j \geq 4$ & 0.064 & 1.1 & 60,000 \\
$\lvert \eta_j \rvert < 2.5$ & 0.063 & 1.1 & 55,000 \\
\hline
$M^{(4j)} \in (3500,4100)$ GeV & 0.027 & 0.55 & 3,300 \\
$M^{(2j)}_{ab},M^{(2j)}_{cd} \in (750,1000)$ GeV & $3.1\times10^{-3}$ & 0.20 & 150 \\
$\cos^2\theta_{\varphi}^* < 0.2$ & $1.3\times10^{-3}$ & 0.13 & 15 \\
$p_{T,[j_1,j_2,j_3,j_4]} > [850,750,550,250]$ GeV & $3.4\times10^{-4}$ & 0.073 & 0.97 \\
\hline
$S/B$ & $3.5\times10^{-4} $ & 0.075 & -- \\
$S/ \sqrt{B}$ ($\mathcal{L} = 300 \textrm{ fb}^{-1}$) & $6.0\times10^{-3}$ & 1.3 & -- \\
$S/ \sqrt{B}$ ($\mathcal{L} =$ 3,000~fb$^{-1}$) & 0.019 & 4.0 & -- \\
Combined $S/ \sqrt{B}$ ($\mathcal{L} =$ 3,000~fb$^{-1}$) & -- & 4.1 & -- \\
\hline \hline
\end{tabular}
\caption{Cut flows of BP1 in SR I ($g_{\rm grav}=5,~g_{g_{\rm KK}}=3.25,~\epsilon=0.75,~ m_{\rm KK}^{(G)}= 4~{\rm TeV},~m_{\rm KK}^{(g)}=3.75~{\rm TeV,~and }~m_\varphi=1~{\rm TeV}$, see also table~\ref{tab:BP}) and QCD 4-jet backgrounds at the 14~TeV LHC with cuts designed for the antler topology. The values of cross-sections are in fb.
The reported no-cut cross-sections in the second and the third columns are the production cross-section of KK graviton multiplied by the associated branching fractions. The pre-selection cuts are $p_{T,[j_1,j_2,j_3,j_4]}> [600,400,200,100]$~GeV, $\Delta R_{jj}>0.4$, and $\lvert \eta_j \rvert<4$.
For the dijet invariant mass window cut, we test whether (at least) a pair of dijet invariant mass values out of three possible pairings satisfy the condition. 
}
\label{Tab:Antler_BP1} 
\vspace{0.5cm}
\begin{tabular}[c]{ l | c | c | c }
\hline \hline
Antler cuts on BP2 in SR I & KK Gluon & Radion & QCD 4 jets \\
\specialrule{.1em}{.05em}{.05em}
No cuts & 0.20 & 3.7 & $1.0\times10^{10}$ \\
Pre-selection cuts & 0.14 & 3.2 & 61,000 \\
$N_j \geq 4$ & 0.14 & 3.2 & 60,000 \\
$\lvert \eta_j \rvert < 2.5$ & 0.14 & 3.1 & 55,000 \\
\hline
$M^{(4j)} \in (3000,3600)$ GeV & 0.063 & 1.6 & 6,400 \\
$M^{(2j)}_{ab},M^{(2j)}_{cd} \in (700,1000)$ GeV & 0.012 & 0.66 & 540 \\
$\cos^2\theta_{\varphi}^* < 0.2$ & $4.8\times10^{-3}$ & 0.43 & 70 \\
$p_{T,[j_1,j_2,j_3,j_4]} > [800,700,500,200]$ GeV & $7.2\times10^{-4}$ & 0.15 & 3.1 \\
\hline
$S/B$ & $2.3\times 10^{-4} $ & 0.049 & -- \\
$S/ \sqrt{B}$ ($\mathcal{L} = 300 \textrm{ fb}^{-1}$) & $7.0\times10^{-3}$ & 1.5 & -- \\
$S/ \sqrt{B}$ ($\mathcal{L} =$ 3,000~fb$^{-1}$) & 0.022 & 4.7 & -- \\
Combined $S/ \sqrt{B}$ ($\mathcal{L} =$ 3,000~fb$^{-1}$) & -- & 4.7 & -- \\
\hline \hline
\end{tabular}
\caption{Cut flows of BP2 in SR I ($g_{\rm grav}=6,~g_{g_{\rm KK}}=4,~\epsilon=1,~ m_{\rm KK}^{(G)}= 3.5~{\rm TeV},~m_{\rm KK}^{(g)}=3.3~{\rm TeV,~and }~m_\varphi=1~{\rm TeV}$, see also table~\ref{tab:BP}) and QCD 4-jet backgrounds. The same explanations including the pre-selection cuts as in the caption of table~\ref{Tab:Antler_BP1} are applicable.
}
\label{Tab:Antler_BP2}
\end{table}

By construction, production cross-section in this signal region is dominated by the antler scenario, so the 2-jet invariant mass conditions described in eq.~\eqref{eq:invmass2body} can capture the associated signal feature while substantially rejecting background events.
We display the unit-normalized $M^{(2j)}$ distributions of BP1 (red), BP2 (blue), and the background (black) in the top-right panel of figure~\ref{fig:obsSR1}, including all six dijet invariant mass values for each event, for illustration. 
As before, the dotted lines mark the input masses of radion (here 1 TeV for both BPs) and we can see that the signal distributions show a peak around $M^{(2j)}=m_\varphi$, as two dijet pairs out of six should give rise to $m_\varphi$.
Based on this observation, we choose our 2-jet invariant mass window cuts for which the detailed numbers are shown in tables~\ref{Tab:Antler_BP1} and \ref{Tab:Antler_BP2}.
The four selected jets allow for three possible pairings of two dijet invariant masses. In our actual selection procedure, we check whether or not there exists at least one pairing in which the two associated dijet invariant mass values lie within the dijet invariant mass window. 

As discussed in section~\ref{sec:simulation}, the angular distribution of radions produced in the KK graviton decay can be used to further suppress the background. We reconstruct the angle $\theta_\varphi^*$ between the momenta of one of the incoming gluons and one of the radions, in the center of mass frame of inital partons. 
The unit-normalized distributions of $\cos^2\theta_\varphi^*$ for BP1 (red), BP2 (blue), and the background (black) are shown in the bottom panel of figure~\ref{fig:obsSR1}. Here the distributions are formed by the events that satisfy all the invariant mass requirements that are discussed above.  
The brown dashed line shows the theoretical expectation for the signal as described in eq.~\eqref{eq:anglespec}, and we find that the signal feature in this angular observable is preserved reasonably well even at the detector level. 
From this exercise, we find that the cut of $\cos^2\theta_\varphi^*<0.2$ is effective in reducing the background, improving the significance by up to a factor of $\sim 2$ as also shown in tables~\ref{Tab:Antler_BP1} and \ref{Tab:Antler_BP2}. 

Finally, we impose cuts on the ordered $p_T$ values of the four hardest jets to further reduce the background. 
We denote the $i$th hardest jet by $p_{T,j_i}$. These cuts are chosen based on the comparison between the distributions of our simulated signal and background events that are shown in figure~\ref{fig:ptordereddist} ($p_{T,j_1}$ in the top-left, $p_{T,j_2}$ in the top-right, $p_{T,j_3}$ in the bottom-left, and $p_{T,j_4}$ in the bottom-right). 
The cuts on $p_{T,j_i}$ are always selected to be harder than the corresponding pre-selection cuts in order to prevent the pre-selection cuts from underestimating background contamination.

The cut flows of BP1 and BP2 in SR I against the QCD 4-jet background are summarized in tables~\ref{Tab:Antler_BP1} and \ref{Tab:Antler_BP2}, respectively. Except for the last four rows, all the entries show cross-sections in the unit of fb according to the cuts imposed. 
The reported no-cut cross-sections in the second and the third columns are the production cross-section of KK gravtion multiplied by the associated branching fractions. 
The pre-selection cuts are given by $p_{T,[j_1,j_2,j_3,j_4]}> [600,400,200,100]$~GeV, $\Delta R_{jj}>0.4$, and $|\eta_j|<4$ for the two benchmark points, while being much softer than the corresponding posterior cuts. 
As mentioned earlier, the cuts are designed for the antler event topology, so one can see that much more signal events from the antler scenario survive after a series of cuts. In both benchmark points the combined statistical significance -- which is defined by $S/\sqrt{B}$ with $S$ and $B$ being the numbers of signal and background events, respectively -- is greater than 4$\sigma$ at an integrated luminosity of 3,000~fb$^{-1}$, so the High-Luminosity (HL) LHC has a great potential to discover the KK graviton arising in the extended warped extra-dimensional model, in the antler topology.  

\subsection{SR II: sequential cascade scenario \label{sec:analysiscascade}}

\begin{figure}[t]
    \centering
   \includegraphics[width=7.5cm]{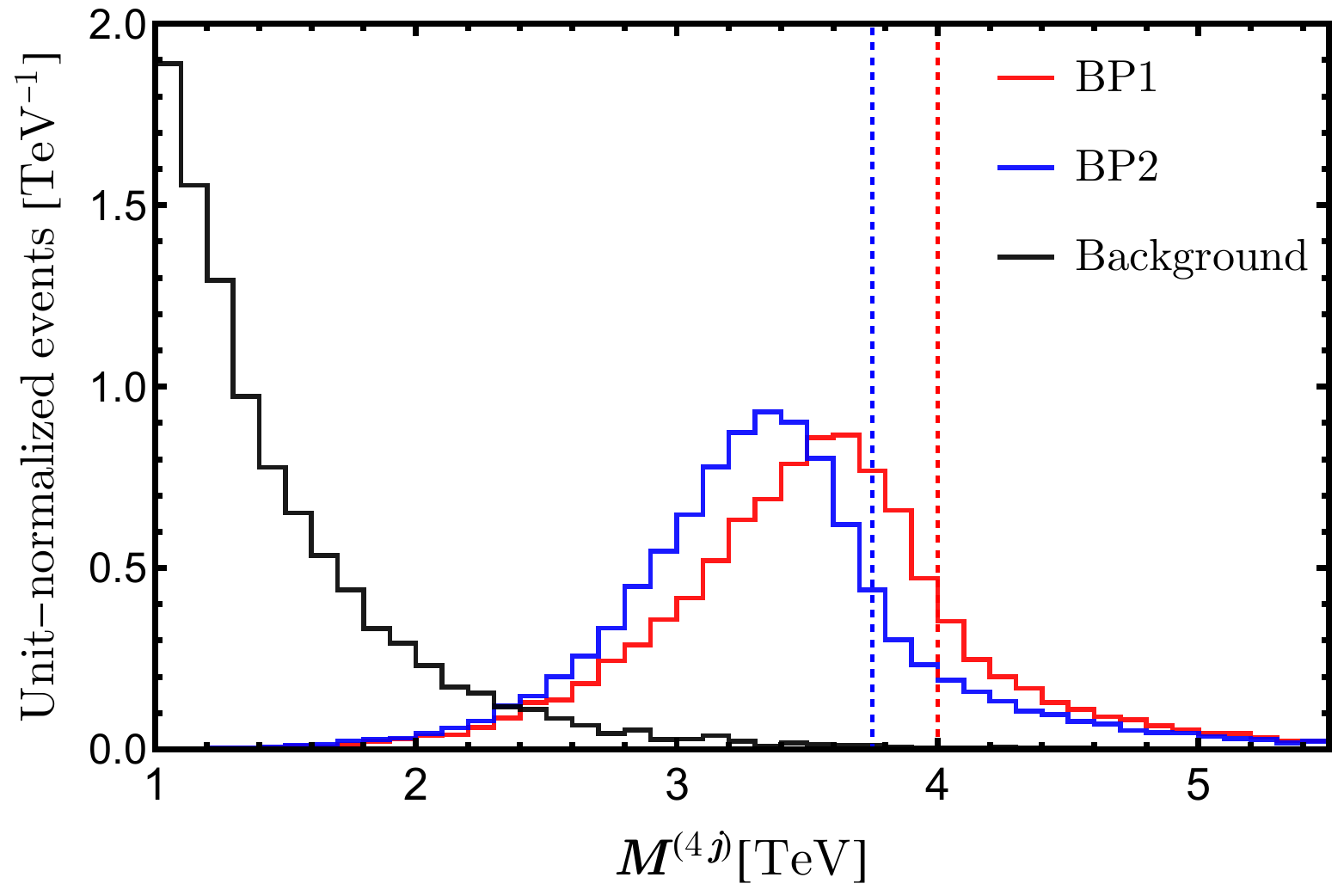}
   \includegraphics[width=7.5cm]{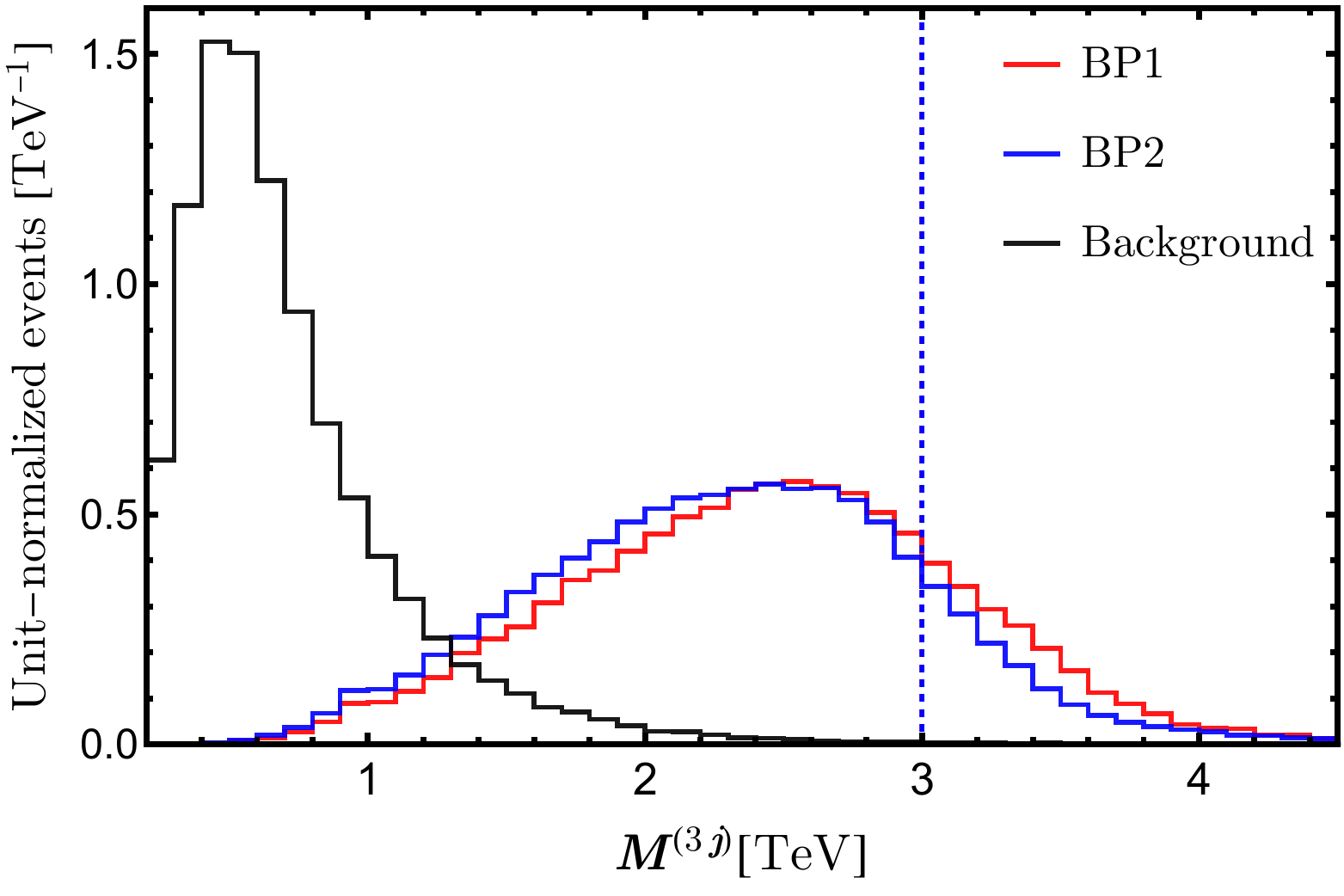} \\
   \includegraphics[width=7.5cm]{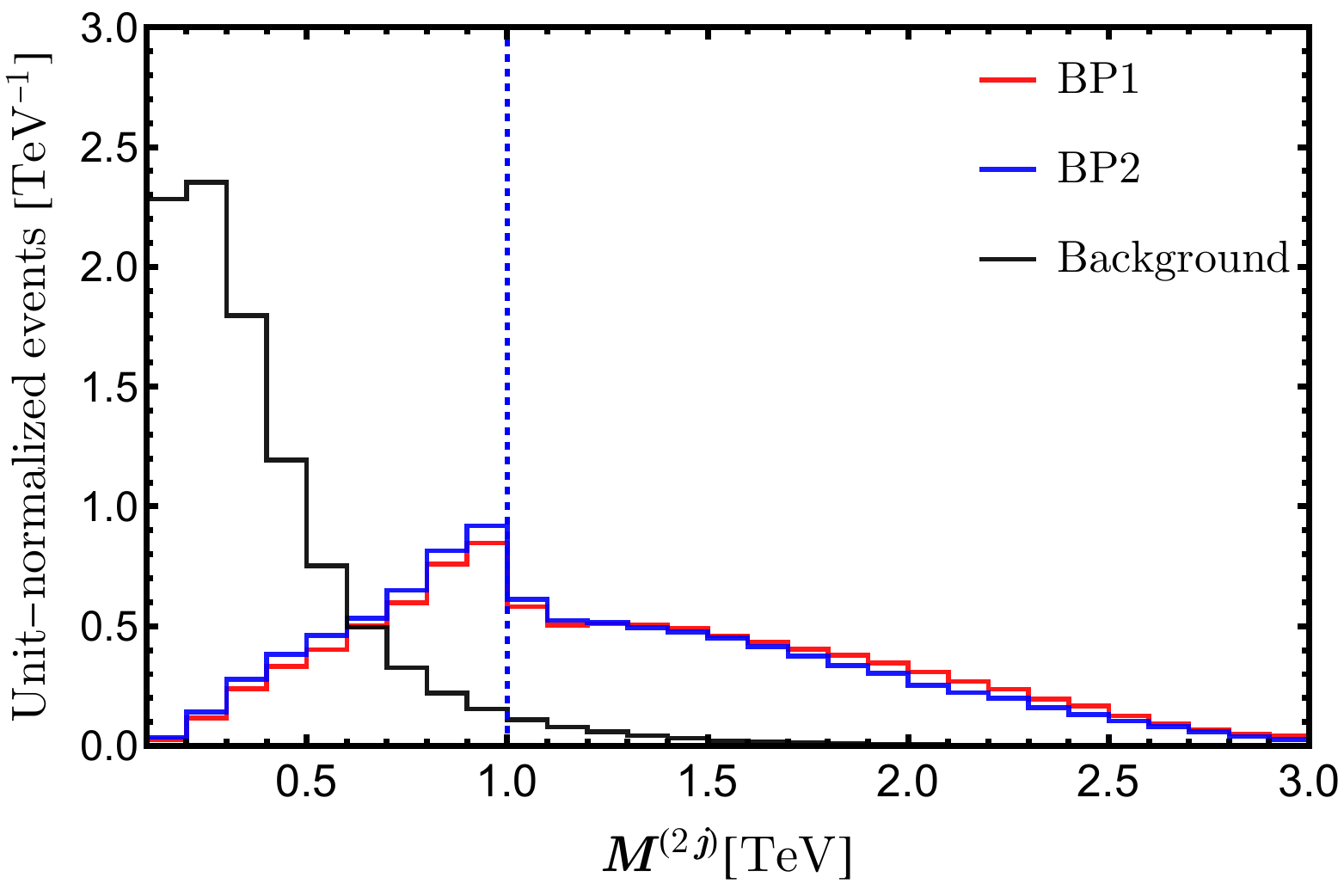}
    \caption{Top left: Unit-normalized $M^{(4j)}$ distributions of BP1 (red) and BP2 (blue) in SR II and the QCD 4-jet background (black). The black dashed line marks the input mass of KK graviton while our 4-jet invariant mass window cuts are indicated by red dashed lines and blue dashed lines for BP1 and BP2, respectively. 
    Top right: Unit-normalized $M^{(3j)}$ distributions of BP1 (red), BP2 (blue), and the background (black).  
    The black dashed line marks the input mass of KK gluon while our trijet invariant mass window cuts are indicated by red dashed lines and blue dashed lines. 
    Bottom: Unit-normalized $M^{(2j)}$ distribution of BP1 (red), BP2 (blue), and the background (black). 
    }
    \label{fig:obsSR2}
\end{figure}

Like the previous antler scenario, the invariant mass variables enable us to reject background events efficiently. First of all, the 4-jet invariant mass distribution contains a resonance peak near the true KK graviton mass. 
Again the four hardest jets in their transverse momentum form an invariant mass value, and the unit-normalized $M^{(4j)}$ distributions of BP1 (red) and BP2 (blue) in SR II and the QCD background (black) are shown in the top-left panel of figure~\ref{fig:obsSR2}. 
The red and blue dotted lines correspond to the input mass values of KK graviton (here 4 TeV and 3.75 TeV, respectively), while the actual peaks of the distributions are located at slightly lower values as before.

Moreover, in each signal event arising from the sequential cascade topology, there should be a 3-jet invariant mass matching the KK gluon mass. Since we do not know the parent resonances associated with each of the four jets, we check whether at least one out of the four possible trijet invariant mass values falls in our 3-jet invariant mass window. Within the selected 3-jet resonance, there should be a dijet (out of three possible pairings) with the associated invariant mass value matching the input radion mass, i.e., we check whether at least a dijet satisfies our 2-jet invariant mass window criterion.   
The top-right panel and the bottom panel of figure~\ref{fig:obsSR2} exhibit the 3-jet invariant mass distributions and the 2-jet invariant mass distributions of BP1 (red) and BP2 (blue) in SR II and QCD background (black), respectively. 
For illustration purpose, we put all four possible trijet invariant mass values in the distribution, whereas for the dijet invariant mass distribution we put all three possible dijet invariant mass values with respect to the three jets forming the KK gluon. 
We clearly see that $M^{(3j)}$ and $M^{(2j)}$ signal distributions show resonance structures near the input KK gluon and radion mass values (dotted lines),\footnote{although the antler topology is sizable in cross-section.} and hence signal events are more densely populated within our invariant mass window choices (see tables~\ref{Tab:Cascade_BP1} and \ref{Tab:Cascade_BP2}) than the backgrounds.   

\begin{figure}[t]
    \centering
    \includegraphics[width=7.5cm]{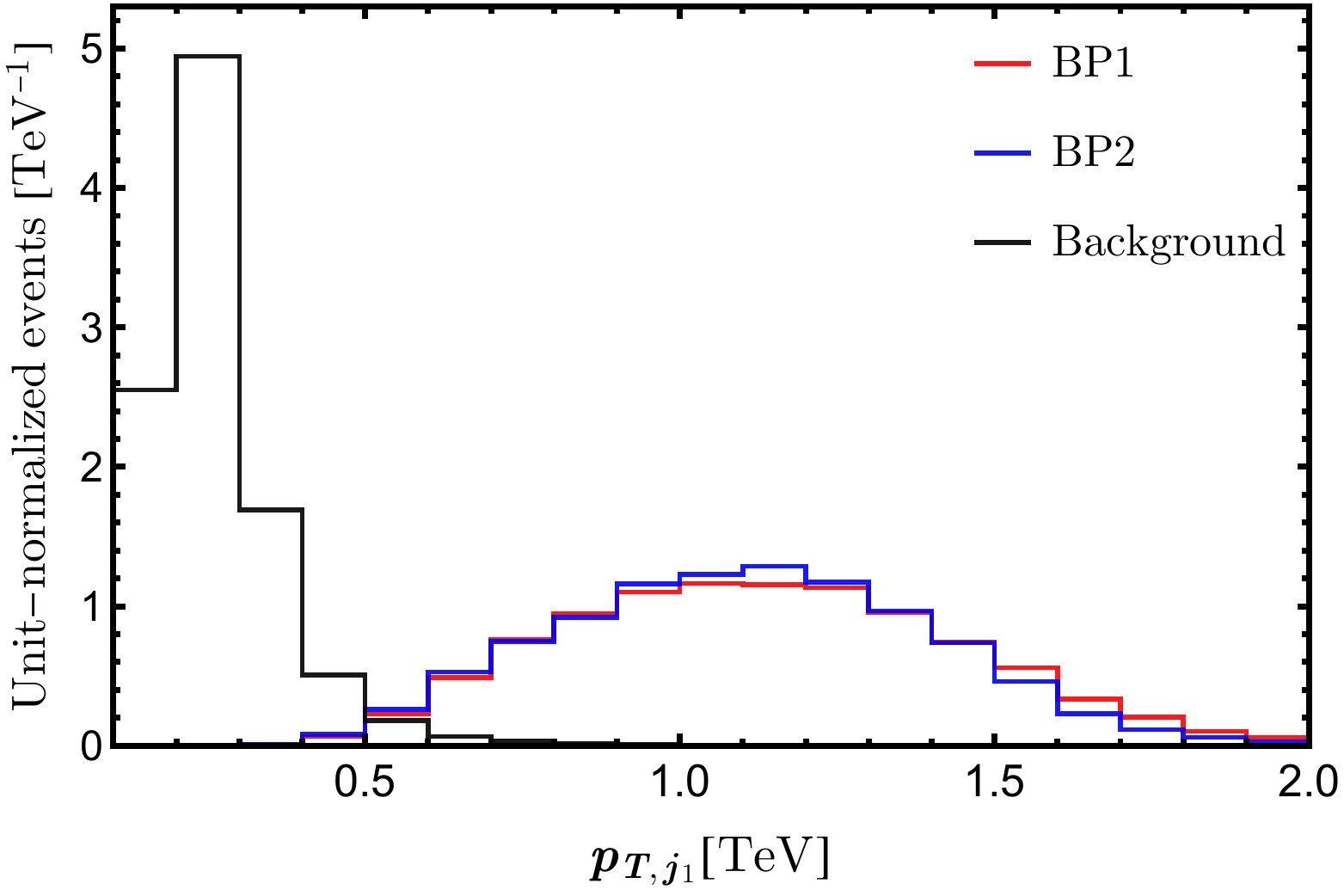}
    \includegraphics[width=7.5cm]{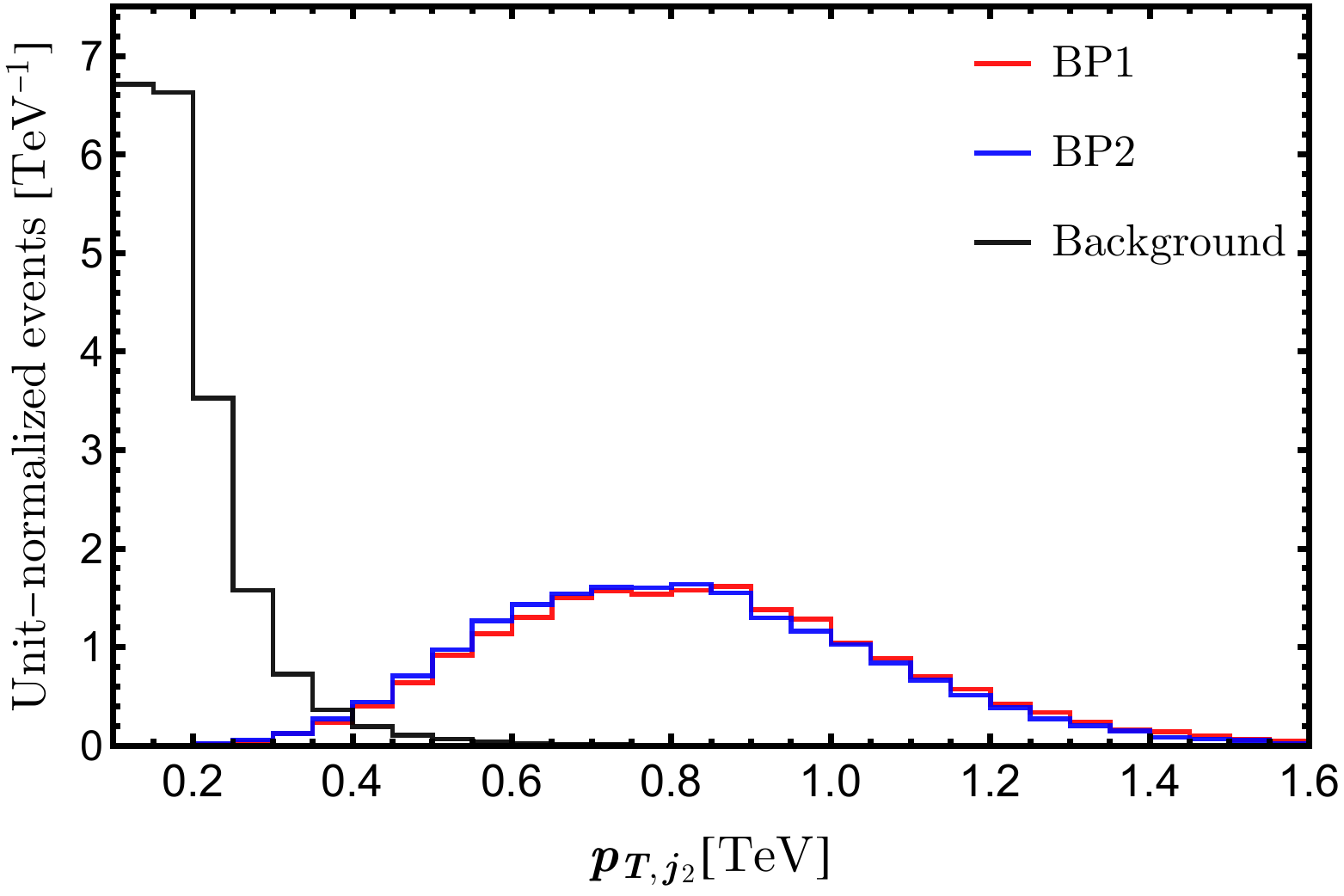} \\
    \includegraphics[width=7.5cm]{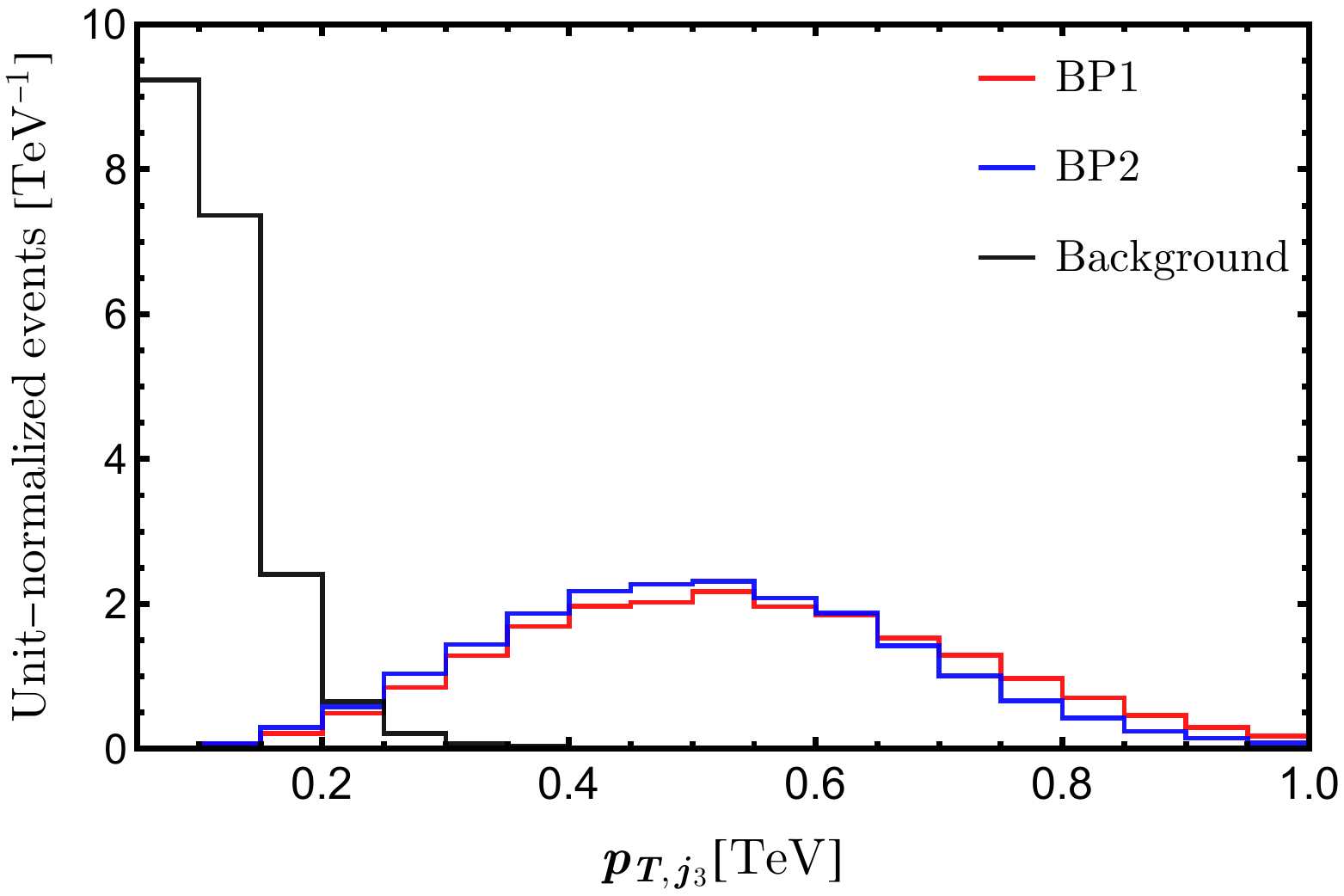}
    \includegraphics[width=7.5cm]{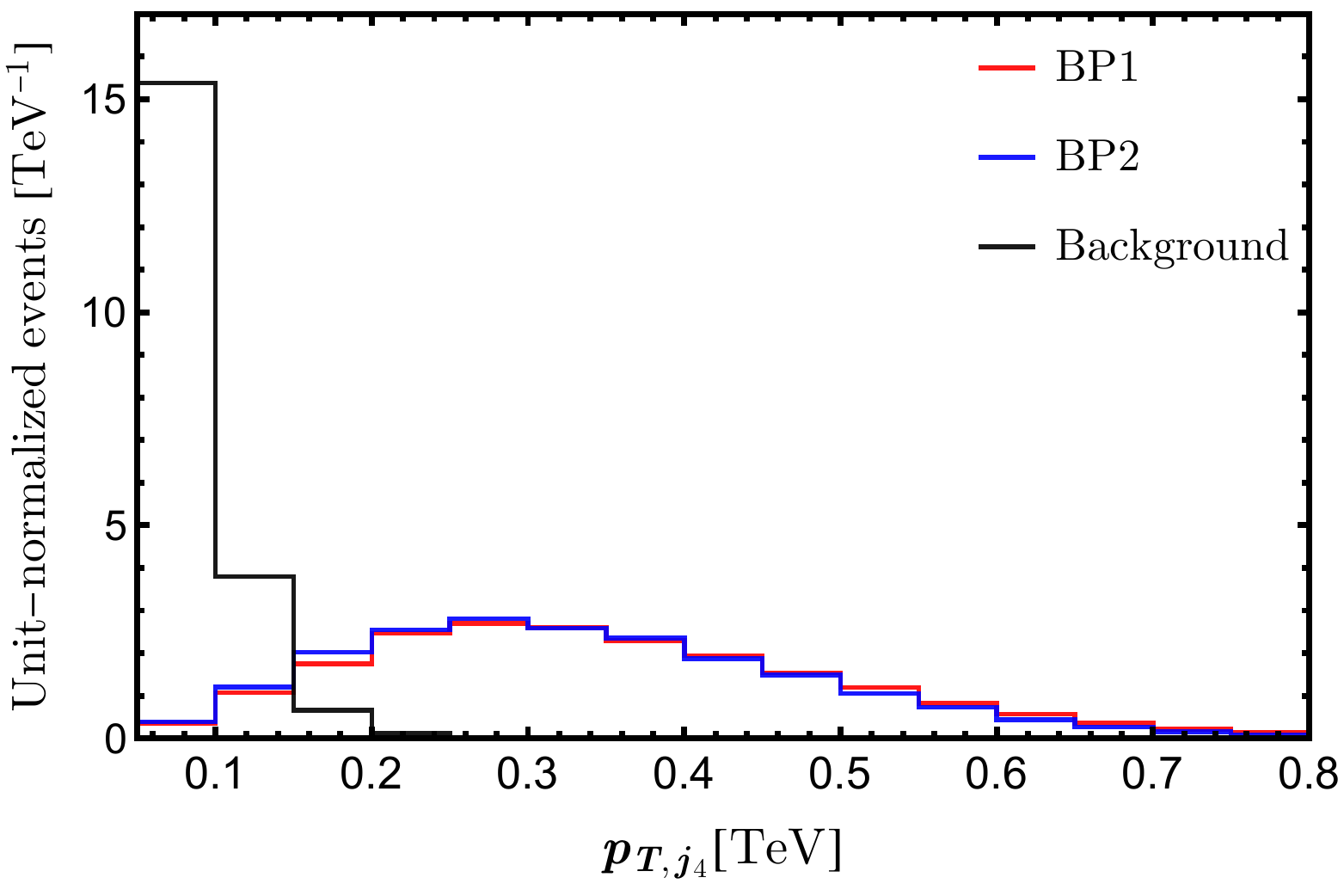}
    \caption{Distributions of ordered transverse momenta, $p_{T,j_i}$ for BP1 (red) and BP2 (blue) in SR II and the QCD 4-jet background (black).} 
    \label{fig:ptordereddistSRII}
\end{figure}

Finally, we apply cuts on the ordered $p_T$ values of the four hardest jets to further suppress the background. The distributions of our simulated signal and background events are shown in figure~\ref{fig:ptordereddistSRII}, providing the idea of choosing the ordered $p_T$ cuts. As in the antler scenario, they are larger than the corresponding pre-selection cuts to avoid underestimating the background. 

The cut flows of BP1 and BP2 in SR II against the QCD 4-jet background are summarized in tables~\ref{Tab:Cascade_BP1} and \ref{Tab:Cascade_BP2}, respectively. All cross-sections are in the unit of fb. 
The reported no-cut cross-sections in the second and the third columns are the production cross-section of KK gravtion multiplied by the associated branching fractions. 
The pre-selection cuts, which are softer than the corresponding analysis cuts, are given by $p_{T,[j_1,j_2,j_3,j_4]}> [600,400,200,100]$~GeV, $\Delta R_{jj}>0.4$, and $|\eta_j|<4$ for the two benchmark points.
The statistical significance is expected to be slightly below 3$\sigma$ for both BPs at an integrated luminosity of 3,000~fb$^{-1}$, so again the HL-LHC can reveal a hint for KK graviton in the cascade topology via an intermediary KK gluon. 

\begin{table}[h!]
\centering
\begin{tabular}[c]{ l | c | c | c }
\hline \hline
Sequential cascade cuts on BP1 in SR II & KK Gluon & Radion & QCD 4 jets \\
\specialrule{.1em}{.05em}{.05em}
No cuts & 1.9 & 0.69 & $1.0\times10^{10}$ \\
Pre-selection cuts & 1.7 & 0.60 & 61,000 \\
$N_j \geq 4$ & 1.7 & 0.60 & 60,000 \\
$\lvert \eta_j \rvert < 2.5$ & 1.7 & 0.59 & 55,000 \\
\hline
$M^{(4j)} \in (3100,4500)$ GeV & 1.2 & 0.47 & 8,600 \\
$M^{(3j)} \in (2100,3300)$ GeV & 1.2 & 0.45 & 7,900 \\
$M^{(2j)} \in (750,1000)$ GeV & 0.80 & 0.35 & 4,100 \\
$p_{T,[j_1,j_2,j_3,j_4]} > [800,700,600,400]$ GeV & 0.14 & 0.10 & 28 \\
\hline
$S/B$ & $5.1\times10^{-3}$ & $3.6\times10^{-3}$ & -- \\
$S/ \sqrt{B}$ ($\mathcal{L} = 300 \textrm{ fb}^{-1}$)& 0.47 & 0.33 & -- \\
$S/ \sqrt{B}$ ($\mathcal{L} =$ 3,000~fb$^{-1}$)& 1.5 & 1.0 & -- \\
Combined $S/ \sqrt{B}$ ($\mathcal{L} =$ 3,000~fb$^{-1}$)& 2.5 & -- & -- \\
\hline \hline
\end{tabular}
\caption{Cut flows of BP1 in SR II ($g_{\rm grav}=6,~g_{g_{\rm KK}}=3.5,~\epsilon=1,~ m_{\rm KK}^{(G)}= 4~{\rm TeV},~m_{\rm KK}^{(g)}=3~{\rm TeV,~and }~m_\varphi=1~{\rm TeV}$, see also table~\ref{tab:BP}) and QCD 4-jet backgrounds at the 14~TeV LHC with cuts designed for the sequential cascade topology.
The values of cross-sections are in fb.
The reported no-cut cross-sections in the second and the third columns are the production cross-section of KK graviton multiplied by the associated branching fractions. The pre-selection cuts are $p_{T,[j_1,j_2,j_3,j_4]}> [600,400,200,100]$~GeV, $\Delta R_{jj}>0.4$, and $|\eta_j|<4$.
For the trijet invariant mass window cut, we test whether (at least) one trijet invariant mass value out of four possible combinations satisfy the condition.
}
\label{Tab:Cascade_BP1} 
\vspace{0.5cm}
\begin{tabular}[c]{ l | c | c | c }
\hline \hline
Sequential cascade cuts on BP2 in SR II & KK Gluon & Radion & QCD 4 jets \\
\specialrule{.1em}{.05em}{.05em}
No cuts & 2.1 & 1.4 & $1.0\times 10^{10}$ \\
Pre-selection cuts  & 1.9 & 1.2 & 61,000 \\
$N_j \geq 4$ & 1.9 & 1.2 & 60,000 \\
$\lvert \eta_j \rvert < 2.5$ & 1.9 & 1.2 & 55,000 \\
\hline
$M^{(4j)} \in (2900,4100)$ GeV & 1.3 & 0.92 & 10,000 \\
$M^{(3j)} \in (2400,3200)$ GeV & 1.2 & 0.85 & 9,400 \\
$M^{(2j)} \in (750,1000)$ GeV & 0.76 & 0.65 & 4,200 \\
$p_{T,[j_1,j_2,j_3,j_4]} > [900,600,500,300]$ GeV & 0.21 & 0.24 & 77 \\
\hline
$S/B$ & $2.8\times10^{-3}$ & $3.2\times10^{-3}$ & -- \\
$S/ \sqrt{B}$ ($\mathcal{L} = 300 \textrm{ fb}^{-1}$)& 0.42 & 0.48 & -- \\
$S/ \sqrt{B}$ ($\mathcal{L} =$ 3,000~fb$^{-1}$)& 1.3 & 1.5 & -- \\
Combined $S/ \sqrt{B}$ ($\mathcal{L} =$ 3,000~fb$^{-1}$)& 2.9 & -- & -- \\
\hline \hline
\end{tabular}
\caption{Cut flows of BP2 in SR II ($g_{\rm grav}=6,~g_{g_{\rm KK}}=3.75,~\epsilon=1,~ m_{\rm KK}^{(G)}= 3.75~{\rm TeV},~m_{\rm KK}^{(g)}=3~{\rm TeV,~and }~m_\varphi=1~{\rm TeV}$, see also table~\ref{tab:BP}) and QCD 4-jet backgrounds. The same explanations including the pre-selection cuts as in the caption of table~\ref{Tab:Cascade_BP1} are applicable.
}
\label{Tab:Cascade_BP2}
\end{table}

Two comments should be made here. First, one can notice that the $M^{(3j)}$ window cut is not as effective as expected in suppressing the backgrounds, in both cases. 
We find that this is rather ``accidental'' in connection with the mass spectra of the two BPs. We can understand this qualitatively as follows. 
The 4-jet invariant mass, ignoring the masses of each jet, can be expressed as
\begin{equation}
    M^{(4j)}=\sqrt{(p_1+p_2+p_3+p_4)^2} \approx \sqrt{6\times 2\langle p_a\cdot p_b\rangle}\,,
\end{equation}
where $p_i$ ($i=1,2,3,4$) denote the jet four momenta and $\langle p_a\cdot p_b\rangle$ stands for the average value of dot products of any jet four-momentum pair, with $a\neq b$. Likewise, the typical or more precisely the root-mean-square (RMS) 3-jet invariant mass can be written as
\begin{equation}
    M^{(3j)}_{\rm RMS} \approx \sqrt{3 \times 2\langle p_a\cdot p_b \rangle }
\end{equation}
If a background event 
passes the cut $M^{(4j)}\approx 4 \; (3.75)$~TeV
as appropriate for BP1 (BP2) in SR II, 
then $\langle p_a\cdot p_b \rangle \approx 1.33 \; (1.17)~{\rm TeV}^2$ resulting in $M^{(3j)}_{\rm RMS} \approx 2.8$ (2.6)~TeV.  
This implies that such a background is likely to {\em already} pass the $M^{(3j)}$ invariant mass window cut
(corresponding to the mass of KK gluon). 
With a different choice for $m_{\rm KK}^{(g)}$, this ineffectiveness would not happen. For example, if the same set of sequential cascade cuts were applied to the BPs in SR I, $M^{(3j)}$ would work as expected, since KK gluon is chosen to be heavier there: of course,
in those cases, BR for KK graviton decay to KK gluon is itself (highly) suppressed due to phase-space.
On the other hand, we see that background events which satisfy $M^{(4j)}\approx 4 \; (3.75)$~TeV have typical $M^{(2j)}_{\rm RMS} = 2 \langle p_a\cdot p_b \rangle \approx 1.6 \; (1.5)$~TeV.
Whereas, for radion mass of 1 TeV that we have chosen for {\em all} the benchmark points
(both in SR I and SR II), the signal events have $M^{(2j)} \approx 1$ TeV.
This rough argument shows that the $M^{(2j)}$ cut is indeed effective in reducing background {\em further}, i.e., beyond the $M^{(4j)}$ one (cf.~the case of $M^{(3j)}$ discussed earlier).

Second, as mentioned earlier, the series of event selection criteria here are designed for the sequential cascade event topology, but we observe that a larger fraction of the radion decay channel events (i.e., $pp\to G_{\rm KK} \to \varphi \varphi$, which inherently have an antler topology instead) satisfy them than the KK gluon events (which do have the sequential cascade topology) in both benchmark points.The reason for this ``counterintuitive'' phenomenon is as follows. In the cascade scenario, there is only one dijet pair forming a radion. Once the dijet pair is selected, there are then two candidate combinations to construct a KK gluon together with each of the two remaining jets. By contrast, in the antler scenario, there are two dijet pairs matching radion. For each of the two dijet pairs, there are two choices to construct a KK gluon, hence in total four possibilities. 
\begin{table}[t]
\centering
\begin{tabular}[c]{ l | c | c | c }
\hline \hline
Antler cuts on BP1 in SR II & KK Gluon & Radion & QCD 4 jets \\
\specialrule{.1em}{.05em}{.05em}
No cuts & 1.9 & 0.69 & $1.0\times 10^{10}$ \\
Pre-selection cuts  & 1.73 & 0.60 & 61,000 \\
$N_j \geq 4$ & 1.73 & 0.60 & 60,000 \\
$\lvert \eta_j \rvert < 2.5$ & 1.69 & 0.59 & 55,000 \\
\hline
$M^{(4j)} \in (3500,4100)$ GeV & 0.62 & 0.27 & 3,300 \\
$M^{(2j)}_{ab},M^{(2j)}_{cd} \in (750,1000)$ GeV & 0.037 & 0.10 & 150 \\
$\cos^2\theta_{\varphi}^* < 0.2$ & 0.015 & 0.068 & 15 \\
$p_{T,[j_1,j_2,j_3,j_4]} > [750,650,550,250]$ GeV & 0.010 & 0.035 & 1.0 \\
\hline
$S/B$ & 0.010 & 0.034 & -- \\
$S/ \sqrt{B}$ ($\mathcal{L} = 300 \textrm{ fb}^{-1}$)& 0.18 & 0.60 & -- \\
$S/ \sqrt{B}$ ($\mathcal{L} =$ 3,000~fb$^{-1}$)& 0.56 & 1.9 & -- \\
Combined $S/ \sqrt{B}$ ($\mathcal{L} =$ 3,000~fb$^{-1}$)& -- & 2.5 & -- \\
\hline \hline
\end{tabular}
\caption{Cut flows of BP1 in SR II ($g_{\rm grav}=6,~g_{g_{\rm KK}}=3.5,~\epsilon=1,~ m_{\rm KK}^{(G)}= 4~{\rm TeV},~m_{\rm KK}^{(g)}=3~{\rm TeV,~and }~m_\varphi=1~{\rm TeV}$, see also table~\ref{tab:BP}) and QCD 4-jet backgrounds, with a set of antler cuts. The same explanations including the pre-selection cuts as in the caption of table~\ref{Tab:Cascade_BP1} are applicable.
}
\label{Tab:Antler_SRIIBP1} \vspace{0.5cm}
\begin{tabular}[c]{ l | c | c | c }
\hline \hline
Antler cuts on BP2 in SR II & KK Gluon & Radion & QCD 4 jets \\
\specialrule{.1em}{.05em}{.05em}
No cuts & 2.1 & 1.4 & $1.0\times 10^{10}$ \\
Pre-selection cuts  & 1.9 & 1.2 & 61,000 \\
$N_j \geq 4$ & 1.9 & 1.2 & 60,000 \\
$\lvert \eta_j \rvert < 2.5$ & 1.9 & 1.2 & 55,000 \\
\hline
$M^{(4j)} \in (3200,3800)$ GeV & 0.79 & 0.60 & 4,900 \\
$M^{(2j)}_{ab},M^{(2j)}_{cd} \in (750,1000)$ GeV & 0.068 & 0.22 & 260 \\
$\cos^2\theta_{\varphi}^* < 0.2$ & 0.032 & 0.14 & 30 \\
$p_{T,[j_1,j_2,j_3,j_4]} > [800,700,500,200]$ GeV & 0.018 & 0.081 & 2.8 \\
\hline
$S/B$ & $6.4\times10^{-3}$ & 0.029 & -- \\
$S/ \sqrt{B}$ ($\mathcal{L} = 300 \textrm{ fb}^{-1}$)& 0.19 & 0.84 & -- \\
$S/ \sqrt{B}$ ($\mathcal{L} =$ 3,000~fb$^{-1}$)& 0.59 & 2.65 & -- \\
Combined $S/ \sqrt{B}$ ($\mathcal{L} =$ 3,000~fb$^{-1}$)& -- & 3.2 & -- \\
\hline \hline
\end{tabular}
\caption{Cut flows of BP2 in SR II ($g_{\rm grav}=6,~g_{g_{\rm KK}}=3.75,~\epsilon=1,~ m_{\rm KK}^{(G)}= 3.75~{\rm TeV},~m_{\rm KK}^{(g)}=3~{\rm TeV,~and }~m_\varphi=1~{\rm TeV}$, see also table~\ref{tab:BP}) and QCD 4-jet backgrounds, with a set of antler cuts. The same explanations including the pre-selection cuts as in the caption of table~\ref{Tab:Cascade_BP1} are applicable.
}
\label{Tab:Antler_SRIIBP2}
\end{table}

At the parton level, $M^{(3j)}$ corresponding to $m_{\rm KK}^{(g)}$ can always be found for the KK gluon events by construction, whereas this is not the case for the events coming from the radion channel unless a $M^{(3j)}$ accidentally takes the value of input $m_{\rm KK}^{(g)}$. However, once detector effects are taken into account and thus rather wide invariant mass windows for $M^{(3j)}$ and $M^{(2j)}$ are introduced, more candidate combinations may increase the chance of passing the two invariant mass window cuts depending on the mass spectrum of the resonances here. Our simulation result suggests this to be the case for our benchmark points. 
This rather high efficiency for radion decay channel events to pass sequential cascade cuts, {\em combined} with their decay BR being significant (even if sub-dominant to KK gluon and SM gluon modes: see the second rows in tables~ \ref{Tab:Cascade_BP1} and \ref{Tab:Cascade_BP2}), implies that the radion decay channel contributes 
substantially to the {\em total} significance with sequential cascade cuts, as seen in tables \ref{Tab:Cascade_BP1} and \ref{Tab:Cascade_BP2}.
On the other hand, a counterpart of this situation does not arise in SR I, i.e., events in the KK gluon decay channel do not have significant efficiency to pass antler cuts, as it is clear from tables~\ref{Tab:Antler_BP1} and \ref{Tab:Antler_BP2}, because the condition of $M_{ab}^{(2j)} \approx M_{cd}^{(2j)}$ is rather hard for the KK gluon decay channel to satisfy. 

Given these two ``accidents'', one may naturally wonder what would happen if an appropriately chosen set of {\em antler} cuts were applied to the two BPs in SR II.  
We perform this exercise and report the corresponding cut flows and the final significance for each BP in tables~\ref{Tab:Antler_SRIIBP1} and \ref{Tab:Antler_SRIIBP2}. 
We see that the cuts designed for the antler event topology yield comparable or slightly improved significance than sequential cascade cuts (see the corresponding significance in tables~\ref{Tab:Cascade_BP1} and \ref{Tab:Cascade_BP2}). 
The reason is (in part) that the cross-section of the radion channel (for
which antler cuts are ideal) is not much smaller than that of the KK gluon channel from the beginning (as indicated above in the context of the second accident), but this happens also because the set of antler cuts is more efficient at reducing SM background than the set of sequential cascade cuts. In particular, the requirement in eq.~\eqref{eq:invmass2body} (i.e., a pair of $M^{(2j)}$ window cuts) is (accidentally) more effective than the requirement in eq.~\eqref{eq:invmass3body} (i.e., $M^{(2j)}$ and $M^{(3j)}$ cuts), as mentioned earlier for the first accident. In other words, even for SR II, the radion decay channel can be the first discovery mode for KK graviton through antler-type cuts, although the KK gluon channel (which has a sequential cascade topology) comes with a larger decay BR.
This implies that evidence for the KK gluon decay channel (via sequential cascade cuts) might be more like an 
interesting cross-check (post-discovery, but rather immediate, given
the significances 
mentioned above) for BPs in SR II, i.e., it is 
useful 
in order to understand the structure of the underlying model. 

\subsection{Comparison with resonant dijet searches \label{sec:analysisdiscussion}}

Finally, we attempt to answer the question whether or not the experimental reach for KK graviton in the 4-jet channels that we have discussed so far is competitive, compared to that of standard dijet searches, as raised at the beginning of this section. 
Since the heavy resonance searches in the dijet channel are well established and the KK graviton in our model has a sizable decay width to a pair of gluons, it is natural to ask the question that we can benefit from our 4-jet channels over the ``standard'' channels, in terms of KK graviton searches. 
We remark that while the BRs for the dijet and the 4-jet final states can be comparable (depending on the exact choice of parameters), the SM background for the latter is expected to be smaller than that for the former, since it arises at higher order in QCD couplings {\it and} is ``forced'' to be even smaller by signal selection criteria to implement more resonance structures (i.e., three invariant mass bumps for the 4-jet signal vs. only one for the dijet signal). Therefore, the proposed 4-jet channels are likely to be the discovery channel (vs. dijet) for the KK graviton in our model. 

Indeed, reference~\cite{Chekanov:2017pnx} performed a dedicated study on the heavy resonance search in the dijet channel at the HL-LHC, assuming that the resonance structure is well approximated by the Gaussian distribution,\footnote{Note that as typical dijet invariant mass distributions are somewhat skewed to the left, this Gaussian approximation might slightly overestimate the bounds. Therefore, the actual bounds could be slightly weaker than estimated.} and presented model-independent experimental sensitivities. 
In table~\ref{tab:compare}, we compare their 95\% C.L. upper limits in resonance production cross-section multiplied by the BR to a jet pair (second row) with the KK graviton production cross-section multiplied by its BR to a pair of jets under our model setup (third row). 
The comparison of these numbers suggests that for all of the benchmark points that we have investigated, the HL-LHC is unlikely to possess sufficient experimental sensitivity to the KK graviton signal in the standard dijet channel. 
By contrast, as summarized in the last row, the searches for KK graviton in the 4-jet channels that we have considered can achieve $\sim 2.5\sigma-4.7\sigma$ significance with the same amount of statistics (i.e., $\mathcal{L}=$ 3,000~fb$^{-1}$). These comparisons clearly justify the heavy resonance search in 4-jet channels, confirming
our above expectation. 

\begin{table}[h]
    \centering
    \begin{tabular}{c|c c c c }
    \hline \hline
         & SR I-BP1 & SR I-BP2 & SR II-BP1 & SR II-BP2  \\
    \specialrule{.1em}{.05em}{.05em}
    95\% C.L. limits in $\sigma\cdot$BR from \cite{Chekanov:2017pnx} & 4.9 fb & 8.0 fb & 4.9 fb & 6.2 fb  \\
    $\sigma(pp\to G_{\rm KK})\cdot {\rm BR}(G_{\rm KK}\to jj)$  & 2.1 fb & 3.5 fb & 0.82 fb & 1.4 fb \\
    \hline
    Significance in the 4-jet channel & \multirow{2}{*}{ 4.1$\sigma$} & \multirow{2}{*}{4.7$\sigma$} & \multirow{2}{*}{2.5$\sigma$} & \multirow{2}{*}{2.9$\sigma$} \\
    with $\mathcal{L}=$ 3,000~fb$^{-1}$ (this work) & &  &  &  \\
    \hline \hline
    \end{tabular}
    \caption{Comparisons between the 95\% C.L. upper limits in resonance production cross-section multiplied by the BR to a jet pair at the HL-LHC ($\mathcal{L}=$ 3,000~fb$^{-1}$)~\cite{Chekanov:2017pnx} (second row) and the KK graviton production cross-section multiplied by its BR to two jets under our model setup (third row).
    For ease of reference we import the expected experimental sensitivities to the KK graviton in the 4-jet channel from tables~\ref{Tab:Antler_BP1} through \ref{Tab:Cascade_BP2} (last row). }
    \label{tab:compare}
\end{table}

\section{Conclusions and Outlook}
\label{sec:conclude}

In this paper, we have analyzed signals at the LHC from the production and decay of KK {\em graviton} in the framework where the SM gauge fields propagate in an {\em extended} 
warped extra dimension.
Akin to  the standard warped model, this framework can address the Planck-weak and flavor hierarchy problems of the SM, while satisfying the EW  and CP/flavor precision bounds
without invoking additional symmetries, albeit then leaving a little hierarchy between the KK scale [$\sim O(10)$ TeV] relevant for cutting-off
Higgs mass divergence and the weak scale. 
Differently though, it still allows KK gauge and graviton modes, along with the  radion, to be accessible to the LHC, i.e., with mass of $O({\rm TeV})$, even though these particles do not quite contribute to curing the problem of Higgs mass divergence.

As is perhaps expected, a common theme for the LHC signals associated with these low-hanging fruits [which provide previews of Higgs compositeness -- as per the AdS/CFT dual picture -- at the slightly higher scale of $O(10)$ TeV] is that we are led to a modification of their decays (as compared to the standard case), whether we consider KK gauge, KK graviton, or radion (these new particles are also of course present in the standard warped model). Namely, the usual suspects for dominant decay modes (i.e., pair of heavy particles of SM) are suppressed, thereby
unmasking {\em pre}-existing, but hitherto sub-dominant, decay channels.
In fact, even more remarkably, this alteration can result in event topologies which thus far have not been probed at the LHC even in the context of other models.
Such a move
then 
motivates new, dedicated searches going forward at the LHC.

In earlier work~\cite{Agashe:2016kfr,Agashe:2017wss,Agashe:2018leo}, this game has been played for gauge KK (in conjunction with radion).
Here, we continued it with the KK graviton, which has a similar tale (and involves KK gluon and radion as well).
Indeed, 
not only are final states from production and decay of KK graviton novel compared to existing searches, but they are also different than those from gauge KK: 
yet more new experimental analyses are thus called for (cf.~standard warped model, where the final states
for KK graviton and gauge signals are similar, hence same searches apply to both particles).
Also, it is noteworthy that we showed that the spin-2 nature of KK graviton can be important in its discovery itself.

In the {\em first} signal region
of our study, we assumed that the KK gluon is close in mass to KK graviton, thus 
decay of KK graviton into KK gluon and SM gluon is negligible.
We then focused on KK graviton decay to a pair of {\em radions}, where each radion, in turn, decays into a pair of gluons, giving a 4-jet final state  with an ``antler'' topology. 
Using cuts designed accordingly (i.e., requiring two distinct pairs of dijet invariant mass bumps and an overall resonance), we have found that there is significant parameter space potentially giving evidence at the LHC (with 3,000~fb$^{-1}$ luminosity) for this radion decay channel of KK graviton, while being consistent with the current bounds on the three relevant new particles (KK graviton, KK gluon, and radion). As concrete examples, we displayed a couple of benchmark points of this sort in this paper, and demonstrated that nearly 5$\sigma$ significance (i.e., discovery) could be achieved for them. Note that there is a substantial decay of KK graviton (directly) into two SM gluons as well, for which the standard dijet search applies, but we showed that this is less sensitive to KK graviton than the 4-jet channel. Hence, it is the decay of KK graviton via radion that is likely to result in the {\em first} discovery of the KK graviton, provided of course that the corresponding new search is undertaken.

Then, we considered the case of KK gluon being significantly lighter than KK graviton such that the {\em KK gluon}, plus SM gluon decay channel for KK graviton becomes significant. 
Indeed, this decay branching ratio (BR) can thus dominate over the decay into a pair of radions or SM gluons.
The KK gluon can decay into SM gluon and radion, followed by the above-mentioned radion decay into two SM gluons,
resulting again in four jets, but now with a 
``sequential cascade'' topology.
We applied cuts suited to it, i.e., a dijet resonance nested inside a 3-jet one and the 4-jet bump.
We realized that the parameter space giving evidence (again, consistent with existing bounds) is more restricted than in the first region above: this is due in part to stringent bounds on the KK gluon from its (other) decay into dijets, cornering us into a mass spectrum
such that the SM background reduction with sequential cascade-type cuts is more difficult than the earlier case of antler cuts. Nonetheless, we showed in this paper a couple of such successful benchmark points whose expected significance could be close to 3$\sigma$. 

Interestingly, 
in this {\em second} signal region of the parameter space, 
it turns out that the radion decay channel of KK graviton also contributes non-negligibly in the above analysis using 
sequential cascade cuts. The reasons are that the radion decay channel has a significant acceptance for the cuts used here, even though they are not tailor-made (i.e., a priori not ideal) for it (this is due to a coincidence related to the selection of our mass spectrum), {\em and} even though the radion decay channel is sub-dominant here in terms of decay BR, it is still non-negligible. In fact, we found (as mentioned above) that the antler topology cuts reduce SM background more (again, due to the choice of mass spectrum) than the sequential cascade. Combining this observation with the still sizable KK graviton decay BR into radions (as indicated above), we find remarkably that the antler-type search (again, where the radion channel is the main contributor) has comparable/slightly more sensitivity (to the KK graviton itself) than the sequential cascade-type search even for the second signal region of the parameter space. 
This implies that once again (i.e., despite 

the KK gluon decay channel of KK graviton,
which has sequential cascade-type topology, having a larger decay BR here), the radion decay channel (via antler-type cuts) can be the first discovery mode for KK graviton (although KK gluon channel is close). 
So, overall, the 
KK gluon decay channel might end up playing the role of 
deciphering the underlying structure of
the new physics (again, for the second signal region).

Just to illustrate, we can then have a 
fascinating discovery story -- for the second signal region
of the parameter space -- as follows here. 
Given that we end up choosing radion, KK gluon masses/couplings in our benchmarks points to be close to those needed to satisfy current bounds, it is clear that (if indeed these points in parameter space are realized in Nature) these particles would be discovered ``soon'', i.e., during the upcoming Run 3 of the LHC. 
Whereas, the discovery of KK graviton (which could happen -- first -- via 4-jet final state, that too the radion channel using antler-type cuts)  would have to ``wait'' till the HL-LHC phase. 
So, digging-out the above KK gluon decay channel of KK graviton using sequential cascade-type cuts 
would be a crucial step in ``connecting'' the two discoveries (again, KK gluon/radion in Run 3 vs. KK graviton with the HL-LHC) which are well separated in time (hence naively looking disparate). 
Again, based on the significances for the two KK graviton channels being comparable, we see that this evidence for the KK gluon channel would appear rather soon after the KK graviton has already been found using the other analysis.

As possible future work, it might be interesting to study the case of radion being (much) lighter than what we assumed here, i.e.,
$\sim O(100)$ GeV instead of $\sim 1$ TeV.
Then the bound from a standard dijet search (i.e., only two well-separated jets) on direct radion production becomes rather weak due to the large background at such low masses.
We have to consider a modified search involving the radion being boosted due to the presence of hard initial state radiation that two gluons from the decay of radion merge, generating a ``fat'' jet: see, for example, ref.~\cite{Sirunyan:2019vxa} for similar searches for a light $Z^{ \prime }$.
Relatedly, the radion produced from the decay of KK gluon and KK graviton (which are much heavier, i.e., a few TeV in mass)
also tends to be boosted, giving a similar di-gluon (fat) jet: 
in fact, the search performed in ref.~\cite{Aad:2020cws} might be relevant in this regard.

Taking a step back from this particular model, we would like to emphasize that our (general) analysis strategy should be applicable to other models which have resonant 4-jet signals.
Indeed, assuming two-body decays at each step of the decay chain, it is easy to see that antler and sequential cascade are the only two topologies for a 4-jet (in general 4-body) final state; both are present in the extended warped model for the KK graviton.
Also, it is possible that in some of the other models, the sequential cascade topology is the way to go even for the first discovery (cf. the parameter space used here for the extended warped model, where antler topology was typically this first discovery mode, with sequential cascade being sort of relegated to the role of post-discovery diagnosis); in fact, this could happen even in our model if we choose flat-ish profiles (in the warped extra dimension) for light SM quarks such that their coupling to KK gluon (which is involved in KK gluon production and decay) is reduced, hence corresponding bound (from dijet search) is relaxed, opening up parameter space of this sort. 

Finally, the big picture here is that it is crucial in general to implement {\em non}-standard search strategies at the LHC (and future colliders), especially because simple alterations to vanilla models of well-established BSM frameworks can lead to novel signals, i.e., it might be essential to cast a wider net in order to catch new physics in its varied forms.

\section*{Acknowledgements}

The authors would like to thank Sergei Chekanov, Hooman Davoudiasl, Peizhi Du, Robert Harris, Zhen Liu, Petar Maksimovic, Benjamin Nachman, Kevin Nash, Raman Sundrum, Nhan Tran, and David Ren-Hwa Yu for discussions.
KA, ME, and DS were 
supported in part by the NSF grant PHY-1914731
and by the Maryland Center for Fundamental Physics.
KA was also supported by the Fermilab Distinguished Scholars Program.
The work of DK is supported by the Department of Energy under Grant No. DE-FG02-13ER41976/DE-SC0009913.

\bibliographystyle{JHEP}
\bibliography{main}

\end{document}